\documentclass[aps,prd,reprint,superscriptaddress]{revtex4-1}

\usepackage{graphicx}         
\usepackage{amssymb}
\usepackage{amsmath}
\usepackage{amsfonts,dsfont}
\usepackage{amsthm}
\usepackage[colorlinks]{hyperref}
\usepackage{color,rotating}
\usepackage{bbm}
\usepackage{array}
\usepackage{pstricks}
\usepackage{pstricks-add}
\usepackage{colonequals}
\usepackage{tikz}
\usepackage{slashed}
\usepackage{bbold}
\usepackage{enumerate}
\usepackage{tensor}

\definecolor{darkgreen}{rgb}{0,0.6,0}
\definecolor{gray}{rgb}{.7,.7,.7}

\DeclareMathAlphabet{\EuRoman}{U}{eur}{m}{n}
\SetMathAlphabet{\EuRoman}{bold}{U}{eur}{b}{n}

 \graphicspath{{./figures/}}

\def\di{\displaystyle}

\def\bg{\begin{eqnarray}\begin{array}{rcl}\displaystyle}
\def\eg{\end{array} &\di    &\di   \end{eqnarray}}
\def\bm#1{\begin{eqnarray}\begin{array}{#1}\di}
\def\bmo#1{\begin{eqnarray*}\begin{array}{#1}\di}
\def\bml#1#2{\begin{eqnarray}\begin{array}{#1}\label{#2}\di}
\def\bgo{\begin{eqnarray*}\begin{array}{rcl}\displaystyle}
\def\ego{\end{array} &\di    &\di \nonumber  \end{eqnarray*}}

\def\btensor#1#2{\renew\left#1\begin{array}{#2}\di}
\def\brtensor#1#2#3{\ren#3\left#1\begin{array}{#2}}
\def\botensor#1#2{\renew\left#1\begin{array}{#2}}
\def\etensor#1{\end{array}\right#1}

\def\eq#1{(\ref{#1})}

\DeclareMathOperator{\Tr}{Tr}

\def\s0#1#2{\mbox{\small{$ \0{#1}{#2} $}}}
\def\0#1#2{\frac{#1}{#2}}

\def\s{\sigma}

\def\ra{\rightarrow}

\def\ren#1{\renewcommand{\arraystretch}{#1}}

\def\renew{\renewcommand{\arraystretch}{1}}

\newcommand{\UV}{{\small UV}}
\newcommand{\IR}{{\small IR}}
\newcommand{\FRG}{{\small FRG}}
\newcommand{\RG}{{\small RG}}
\newcommand{\TT}{\text{\tiny{TT}}}
\newcommand{\TText}{{\small TT}}
\newcommand{\QCD}{{\small QCD}}

\newcommand{\DEP}{{\small DEP}}

\allowdisplaybreaks

\begin{document}
\title{Asymptotic safety of gravity-matter systems}

%
\author{J. Meibohm}
\affiliation{Institut f\"ur Theoretische Physik, Universit\"at Heidelberg,
Philosophenweg 16, 69120 Heidelberg, Germany}
\author{J. M. Pawlowski}
\affiliation{Institut f\"ur Theoretische Physik, Universit\"at Heidelberg,
Philosophenweg 16, 69120 Heidelberg, Germany}
\affiliation{ExtreMe Matter Institute EMMI, GSI Helmholtzzentrum f\"ur
Schwerionenforschung mbH, Planckstr.\ 1, 64291 Darmstadt, Germany}
\author{M. Reichert}
\affiliation{Institut f\"ur Theoretische Physik, Universit\"at Heidelberg,
Philosophenweg 16, 69120 Heidelberg, Germany}
                                  
%
\begin{abstract}
  We study the ultraviolet stability of gravity-matter systems for
  general numbers of minimally coupled scalars and fermions. This is
  done within the functional renormalisation group setup put
  forward in \cite{Christiansen:2015rva} for pure gravity. It includes
  full dynamical propagators and a genuine dynamical Newton's
  coupling, which is extracted from the graviton three-point function. 
  
  We find ultraviolet stability of general gravity-fermion
  systems. Gravity-scalar systems are also found to be ultraviolet
  stable within validity bounds for the chosen generic class of
  regulators, based on the size of the anomalous
  dimension. Remarkably, the ultraviolet fixed points for the
  dynamical couplings are found to be significantly different from
  those of their associated background counterparts, once matter
  fields are included. In summary, the asymptotic safety scenario does
  not put constraints on the matter content of the theory within the
  validity bounds for the chosen generic class of regulators.

\end{abstract}

\maketitle

\section{Introduction}
The asymptotic safety scenario proposed by S.\ Weinberg
\cite{Weinberg:1980gg} almost 40 years ago has received growing
attention in the last decades. It provides a promising route towards
the formulation of quantum gravity as a non-perturbatively
renormalisable quantum field theory of the metric.  In terms of the
renormalisation group, the asymptotic safety scenario conjectures the
existence of a non-trivial ultraviolet (\UV) fixed point of the
renormalisation group flow.

The development of modern functional renormalisation group (\FRG)
techniques and their application to quantum gravity
\cite{Reuter:1996cp,Wetterich:1992yh} has led to strong evidence for
the non-trivial \UV{} fixed point for pure gravity. It was
first found in basic Einstein-Hilbert approximations
\cite{Reuter:1996cp,Souma:1999at,Reuter:2001ag} and later confirmed in
more elaborate truncations
\cite{Christiansen:2012rx,Christiansen:2014raa,Christiansen:2015rva,Donkin:2012ud,
  Falls:2014tra,Lauscher:2002sq,Codello:2006in,Codello:2007bd,Codello:2008vh,%
  Machado:2007ea,Benedetti:2009rx,Eichhorn:2009ah,Manrique:2011jc,Rechenberger:2012pm,%
  Codello:2013fpa,Falls:2013bv,Falls:2014zba,Falls:2015qga,Gies:2015tca}, for reviews
see
\cite{Niedermaier:2006wt,Percacci:2007sz,Litim:2011cp,Reuter:2012id}.

One of the most interesting open physics questions concerns the \UV{}
completions of the Standard Model of particle physics.  This requires
the investigation of the \UV{} stability of interacting gravity-matter
systems, and in particular those with large numbers of matter
fields. First interesting results and developments in this direction
have been obtained in
\cite{Dou:1997fg,Percacci:2002ie,Percacci:2003jz,Folkerts:2011jz,
  Dona:2012am,Dona:2013qba,Dona:2014pla,Oda:2015sma}. An interacting
fixed point in gravity-matter systems requires a non-trivial interplay
of the fluctuation dynamics of all involved fields. In other theories
it is well-known, that the inclusion of additional fields may change
the nature of the theory. For example, in \QCD{} with many quark
flavours asymptotic freedom is lost, thus rendering the \UV-limit of
the theory ill-defined for a large number of quarks. Analogously,
matter fields could potentially spoil asymptotic safety in combined
systems of gravity and matter. This has indeed been observed in
\cite{Percacci:2002ie} in the background field approximation and in
\cite{Dona:2013qba,Dona:2014pla} with a mixed approach, where the
background field approximation for the couplings is augmented with
dynamical anomalous dimensions. In the background field approximation
no distinction is made between dynamical and background
fields. However, the differences between these fields are potentially
of qualitative nature, see
\cite{Pawlowski:2002eb,Litim:2002ce,Litim:2002hj,Pawlowski:2003sk,
  Pawlowski:2005xe,Folkerts:2011jz}.  More recently, a more careful
treatment of background and dynamical fluctuating fields has been
provided by, e.g., the \FRG{} setup with dynamical correlation
functions in
\cite{Christiansen:2012rx,Christiansen:2014raa,Christiansen:2015rva},
the use of the geometrical effective action and the corresponding
Nielsen identities
\cite{Pawlowski:2003sk,Branchina:2003ek,Pawlowski:2005xe,Donkin:2012ud,%
  Demmel:2014hla,Demmel:2015zfa,Safari:2015dva}, or bi-metric
approaches \cite{Manrique:2009uh,Manrique:2010am}.

In this work we analyse the influence of scalar and fermionic matter
on the non-trivial \UV{} fixed point of quantum gravity in the
dynamical \FRG{} setup put forward in \cite{Christiansen:2015rva}. The
matter contributions to the quantum gravity system are extracted, for
the first time, from the higher order dynamical correlation functions
in the framework of the \FRG{}. As introduced in
\cite{Christiansen:2012rx,Christiansen:2014raa,Christiansen:2015rva}
we analyse a system of vertex flows evaluated at flat Euclidean
background. We also introduce a validity bound on the generic class of
regulators used here, based on the size of the anomalous
dimensions. This regime includes an arbitrary numbers of fermions,
whereas it restricts the number of allowed scalars that can be
discussed with the present generic class of regulators to a maximum
$\lesssim 20$. Within this regime of validity we find that the
\UV-fixed point persists and remains \UV{} stable. We also find that
the \UV{} fixed points for the dynamical couplings are significantly
different from those of their associated background counterparts, once
matter fields are included. In summary, the asymptotic safety scenario
does not put constraints on the matter content of the theory within
the validity bounds for the chosen generic class of regulators.

\section{Functional Renormalisation Group}\label{sec:frg}
The basic quantity in the functional renormalisation group approach
\cite{Wetterich:1992yh,Reuter:1993kw,Ellwanger:1993mw,Morris:1993qb} is the quantum effective action
$\Gamma[\bar g,\phi]$, where $\phi$ is a superfield containing the
dynamical fields of the theory and $\bar g_{\mu\nu}$ is the background
metric. For our set of fields, $\phi$ reads
\begin{align}
	\phi = (h,c,\bar c,\psi_i,\bar \psi_j, \varphi_l)\,,
\end{align}
where $h_{\alpha \beta}$ and $(\bar c_\mu, c_\nu)$ are the fluctuating
graviton and the (anti-) ghost fields, respectively. The fermion
fields $(\bar \psi_i,\psi_j)$, carrying the flavour indices
$i,j\in1\ldots N_f$, and the real scalars $\varphi_l$
of flavour $l=1\ldots N_s$ constitute the matter contributions to
$\phi$.

\begin{figure}[t]
  \includegraphics[width=8.5cm]{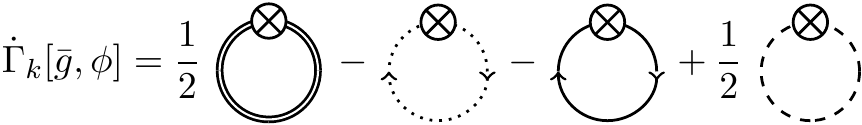}
  \caption{Flow equation for the scale dependent effective action
    $\Gamma_k$ in diagrammatic representation. The double, dotted,
    solid and dashed lines correspond to the graviton, ghost, fermion
    and scalar propagators, respectively. The crossed circles denote
    the respective regulator insertions.}
  \label{fig:flow}
\end{figure}

The scale dependent effective action $\Gamma_k[\bar g,\phi]$ is
formally defined by introducing $k$-dependent \IR{} regulators
$R_k^\phi$ for the fluctuation fields $\phi$ on the level of the path
integral. We call the scale parameter $k$ the renormalisation
scale. The regulators are quadratic in the fluctuating fields, which
requires to introduce a background metric $\bar g_{\mu\nu}$ for metric
theories of gravity.  The physical (full) metric $g_{\mu \nu}$ is
given by a linear split between background metric $\bar g_{\mu\nu}$
and the fluctuation field $h_{\mu\nu}$ according to $g_{\mu\nu} = \bar
g_{\mu\nu}+ h_{\mu\nu}$.  The scale dependent effective action
$\Gamma_k[\bar g,\phi]$ obeys a one-loop flow equation.  For the given
field content $\phi$ the latter reads
\begin{align}\label{eq:flow}
  \dot \Gamma_k =& \012 \Tr\left[ \01{\Gamma_k^{(2)}+R_{k}}\dot R_{k}
  \right]_{hh}\hspace*{-0.3cm}
  - \Tr\left[ \01{\Gamma_k^{(2)}+R_{k}}\dot R_{k}\right]_{\bar c c} \notag \\
  -&\Tr\left[\01{\Gamma_k^{(2)}+R_{k}}\dot R_{k}\right]_{\bar \psi
    \psi}\hspace*{-0.3cm}+\012 \Tr\left[\01{\Gamma_k^{(2)}+R_{k}} \dot
    R_{k}\right]_{\phi\phi}\hspace*{-0.15cm}.
\end{align}
Here we have introduced the notation $\dot
f=\partial_t f$ where $t=\ln(\0k{k_0})$ is the renormalisation time 
with some reference scale $k_0$. \autoref{fig:flow} depicts
equation \eqref{eq:flow} in terms of diagrams.
Since \eqref{eq:flow} is not solvable in general, $\Gamma_k[\bar
g,\phi]$ is truncated to a finite dimensional set of
operators. In our approach, the latter set is given
  by the $n$-point correlation functions $\Gamma_k^{(hh)}$,
  $\Gamma_k^{(hhh)}$, $\Gamma_k^{(\bar\psi \psi)}$ and
  $\Gamma_k^{(\phi \phi)}$, where we employ the condensed notation for
  the $k$-dependent 1\,{\small PI} $n$-point functions
\begin{align}
  \Gamma_k^{(\phi_1\ldots\phi_n)}[\bar g, \phi] := \0{\delta^n
    \Gamma_k[\bar g,\phi]}{\delta \phi_1 \ldots \delta \phi_n}\,.
\end{align}
The flow equations for this set of operators are obtained by taking
field variations of the flow equation \eqref{eq:flow} and expanding
the full scale dependent effective action in powers of the fields
according to
\begin{align}\label{eq:vexp}
  \Gamma_k[\bar g, \phi] = &
  \sum_{n=0}^{\infty}\frac{1}{n!}\Gamma^{(\phi_1\ldots\phi_n)}_k[\bar
  g,0]\phi_1\cdots\phi_n \notag \\
  = & \, \Gamma_k[\bar{g},0] + \Gamma^{(h)}_{k}[\bar{g},0] h
  + \012 \Gamma^{(2h)}_{k}[\bar{g},0] h^2  \notag \\
  & + \01{3!}\Gamma^{(3h)}_{k}[\bar{g},0] h^3
  + \012 \Gamma^{(\overline{c}c)}_{k}[\bar{g},0] \overline{c} \, c   \notag \\
  & + \012 \Gamma^{(\overline{\psi}\psi)}_{k}[\bar{g},0]
  \overline{\psi} \, \psi + \012
  \Gamma^{(\varphi\varphi)}_{k}[\bar{g},0] \varphi^2 + \ldots \, .
\end{align}
This vertex expansion of the scale dependent effective action was
introduced in
\cite{Christiansen:2012rx,Christiansen:2014raa,Christiansen:2015rva}
in the context of pure quantum gravity. In other related works,
anomalous dimensions were computed with vertex expansions on a flat
background and were used in combination with the background-field
approach \cite{Eichhorn:2010tb,Codello:2013fpa,Dona:2013qba}.
Together with \cite{Christiansen:2015rva}, however, the present work
is the first minimally self-consistent analysis of such vertex flows
in quantum gravity.

Note that $\Gamma_k[\bar g,\phi]$ in \eqref{eq:vexp} is expanded about
an, a priori, arbitrary fixed metric background $\bar g_{\mu\nu}$. As
we will see, however, the present setup allows us to evaluate all
relevant flow parameters on a flat Euclidean background, i.e.\ $\bar
g_{\mu\nu}=\delta_{\mu\nu}$.  In \eqref{eq:vexp}, the zero-point
function $\Gamma_k[\bar{g},0]$ and the one-point function
$\Gamma^{(h)}_{k}[\bar{g},0]$ are non-dynamical (background-)
quantities that do not feed back into the flow of the dynamical
$n$-point functions.  Therefore we first focus on the computation of
the latter ones and afterwards, in
\autoref{sec:background-field-approx}, use the solution of the
dynamical couplings for a self-consistent computation of the
background couplings.  Since the right hand side of the flow equation
\eqref{eq:flow} contains second variations of the fields, the flows
for the respective $n$-point functions contain $n$-point vertices up
to order $n+2$. More precisely, the present setup requires the
evaluation of vertices with up to five fields.

We also want to briefly compare the present expansion scheme with the
standard heat kernel expansion in the background field
approximation. In this approximation it is assumed that the scale dependent
effective action is a functional of only one single metric field $g=\bar
g+h$. Note that this approximation has the seeming benefit of a
diffeomorphism invariant expansion scheme and a closed, diffeomorphism
invariant effective action. However, the background field
approximation does not satisfy the non-trivial Slavnov-Taylor
identities for the dynamical metric $h$ as well as the Nielsen
identity, that link $\bar g$-dependences and $h$-dependences, see in
particular
\cite{Pawlowski:2003sk,Pawlowski:2005xe,Folkerts:2011jz,Donkin:2012ud,Demmel:2014hla, Demmel:2015zfa, Safari:2015dva}.
Hence, while based on a diffeomorphism-invariant effective action, the
background field approximation is at odds with diffeomorphism
invariance for this very reason. Note that this also implies that
background independence is at stake. The potential severeness of the
related problems has been illustrated early on at the simpler example
of a non-Abelian gauge theory in \cite{Litim:2002ce}.
These problems can either be resolved 
in the present approach within a flat background expansion, the geometrical effective action approach, see 
\cite{Branchina:2003ek,Pawlowski:2003sk,Pawlowski:2005xe,Donkin:2012ud}. 
or in the bi-metric approach, see
\cite{Manrique:2009uh,Manrique:2010am}. Results within these
approaches also allow for a systematic check of the reliability of the
background field approximation.  Note also that the full resolution of
the background independcence within the bi-metric approach requires
the computation of $h$-correlation function functions of the order two
and higher as it is only these correlation functions that enter the
flow equation on the right hand side. So far, this has not been
undertaken.

The heat kernel computation expands the solution in powers of the
Ricci scalar $R$, to wit
\begin{align}
  \dot \Gamma_k[g] = c_0 \int d^4x\, \sqrt{g} + c_1 \int d^4x\,
  \sqrt{g}\,R + \mathcal{O}(R^2) \,.
\end{align}
The coefficients $c_0=c_0 (\dot {\bar g},\dot {\bar \lambda})$ and
$c_1=c_1 (\dot {\bar g},\dot {\bar \lambda})$ are related to flow of
the background couplings.  By computing the flow for the graviton
two-point function for this hypothetical situation according to
\begin{align}
  \mathcal{F}\circ \dot \Gamma_k^{(2h)} [\bar g ] \Big|_{\bar g =
    \delta} = c_0 \mathcal{T}^{(2h)}(0) + c_1
  \mathcal{T}^{(2h)}(\mathbf{p})\,,
\end{align}
where $\mathcal{F}$ denotes the Fourier transform, we observe that the
coefficients $c_0$ and $c_1$ are obtained analogously from the
momentum independent and momentum dependent parts of the graviton
two-point function, respectively.  The tensor structures $\mathcal{T}$
are defined later in \eqref{eq:Tensorstructures}.  In consequence, we
extract exactly the same information from the flow within the flat
vertex expansion that is obtained in the heat kernel approach.  In
case of higher order operators, we are even able to distinguish
between the flows of e.g.\ $R^2$ and $R_{\mu\nu}R^{\mu\nu}$.
Considering the realistic situation that the flow is not a functional
of only one single metric but of a background and a fluctuating field,
the vertex expansion further conveniently disentangles the flows of
their corresponding couplings.  In summary the present approach
retains the results of standard heat kernel computation although it is
evaluated on a flat background but has significant advantages in the
non-single metric of quantum gravity.

\begin{figure*}
	\includegraphics[width=\linewidth]{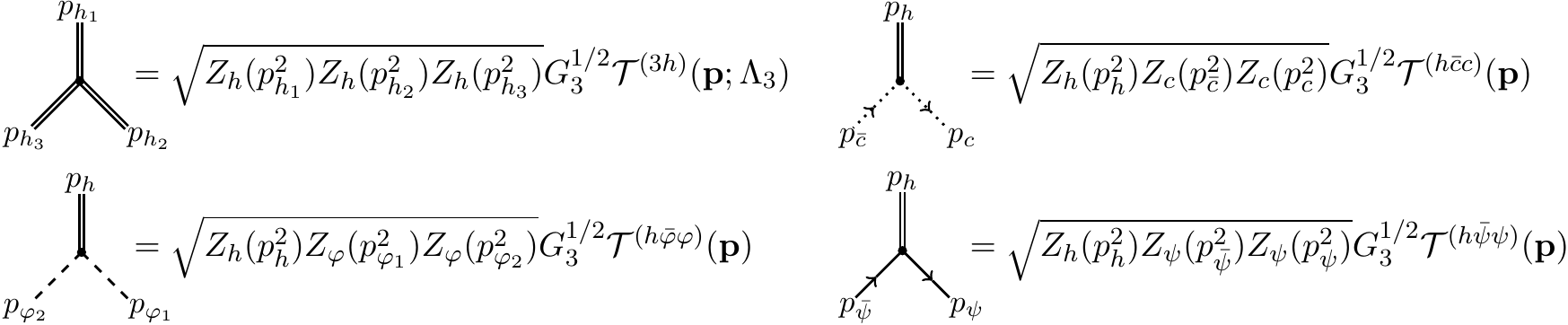}
	\caption{Vertex dressing of all three-point vertices used in
          this work. The vertex dressing consist of the respective
          wave function renormalisations, couplings and tensor
          structures. The first line in the figure depicts all pure
          gravity three-point vertices while the second line shows the
          ones with gravity-matter-interactions.}
	\label{fig:vertices}
\end{figure*}

In order to obtain running couplings from the flow of the $n$-point
functions we employ a vertex dressing according to
\begin{align}\label{eq:vert}
  \Gamma_k^{(\phi_1\ldots\phi_n)}=\sqrt{\prod_{i=1}^n
    Z_{\phi_i}(p_i^2)}\,G_{n}^{\0n2-1}
  \mathcal{T}^{(\phi_1\ldots\phi_n)}\,,
\end{align}
where $Z_{\phi_i}$ denote the wavefunction renormalisations of the
respective fields in $\phi$ which are functions of the field momenta
$p^2_i$. $\mathcal{T}^{(\phi_1\ldots\phi_n)}$ is the tensor structure
of the respective vertex and shall be defined in
\eqref{eq:Tensorstructures}.  In general, we assign to any $n$-vertex
an individual, momentum dependent Newton's constant
$G_{n}(\mathbf{p})$, with $\mathbf{p}=(p_1,\dots,p_n)$. In this work,
however, we approximate all $G_n$ as one, momentum-independent
coupling, $G_n(\mathbf{p}) \equiv G_3 =: G$. Note, that $Z_\phi$ and
$G$ are scale dependent, although we drop the subscript $k$ here and
in the following for notational convenience.  In
\autoref{fig:vertices} the vertex dressing of all involved the
three-point vertices are given according to \eqref{eq:vert}.
Generalisations to higher order vertices can be inferred from
\eqref{eq:vert}.
Note, that \eqref{eq:vert} suggests an expansion in rescaled fields
$\bar\phi$ and rescaled vertices $\bar \Gamma^{(\phi_1\ldots\phi_n)}$
with
\begin{align}
  \phi = \0{\bar \phi}{ \sqrt{Z_\phi}}\,, \qquad
  \bar\Gamma^{(\phi_1\ldots\phi_n)} =
  \0{\Gamma^{(\phi_1\ldots\phi_n)}}{\sqrt{\prod_{i=1}^n
      Z_{\phi_i}(p_i^2)}} \simeq G_n^{\0n2-1} \,,
\end{align} 
see also \cite{Fischer:2009tn,Pawlowski:2005xe,Christiansen:2014raa}. Such a rescaling
absorbs the \RG-running of the vertices in the fields, and hence is an
expansion in RG-invariant, but cutoff-dependent, quantities, for more
details on this aspect see \cite{Fischer:2009tn,Pawlowski:2005xe,Christiansen:2014raa}.
The underlying structure is elucidated by the kinetic term
$\bar\Gamma^{(\phi_1\phi_2)}$: it has the classical form without wave
function renormalisation, and hence does not scale under
RG-transformations.  This discussion highlights the r$\hat{\rm o}$le
of the couplings $G_n$ as RG-invariant running couplings.

The tensor structures $\mathcal{T}$ are given by
variations of the classical action $S$ with respect to the fluctuation
fields. More precisely, the latter read
\begin{align}\label{eq:Tensorstructures}
  \mathcal{T}^{(\phi_1\ldots\phi_n)}(\mathbf{p}; \Lambda_{n}) =
  S^{(\phi_1\ldots\phi_n)}(\mathbf{p};\Lambda \rightarrow
  \Lambda_{n},G_N \rightarrow 1)\,.
\end{align}
In \eqref{eq:Tensorstructures} the classical action $S$ is given by
the Einstein-Hilbert action added by covariant fermion and scalar
kinetic terms according to
\begin{align}\label{eq:EH}
  S = S_\text{EH}+ \int \mathrm d^4x \sqrt{g} \bar \psi_i\slashed
  \nabla\psi_i+ \012\int \mathrm d^4x \sqrt{g} g_{\mu\nu}\partial^\mu
  \varphi_l \partial^\nu\varphi_l\,,
\end{align}
where we used the conventional slash-notation for the contraction of
the spin-covariant derivative $\nabla^\mu$ with gamma matrices. The
covariant kinetic terms for the matter fields in \eqref{eq:EH} lead to
minimal coupling between gravity and matter in the present
truncation. For the formulation of fermions in curved spacetime we use
the the spin-base invariance formalism introduced in
\cite{Weldon:2000fr,Gies:2013noa,Lippoldt:2015cea}. This allows to
circumvent possible ambiguities arising in the vielbein formalism and
relies on spacetime dependent $\gamma$-matrices and the
spin-connection $\Gamma^\mu$. As a result, $\slashed \nabla$ reads
\begin{align}
  \slashed \nabla =
  g_{\mu\nu}\gamma(x)^\mu\nabla^\nu=g_{\mu\nu}\gamma(x)^\mu(\partial^\nu
  +\Gamma(x)^\nu)\,,
\end{align}
if it acts on a spinor as in \eqref{eq:EH}. In the following, we drop
the explicit spacetime dependence of the latter quantities for a more 
convenient notation. The gauge-fixed Einstein-Hilbert action $S_\text{EH}$ 
in \eqref{eq:EH} reads
\begin{align}
  S_\text{EH}=\01{16\pi G_N}\int\mathrm d^4x
  \sqrt{g}(2\Lambda-R)+S_\text{gf}+S_\text{gh}\,,
\end{align}
where $\Lambda$ denotes the classical cosmological constant and $R$ is
the curvature scalar. The terms $S_\text{gf}$ and $S_\text{gh}$ are
the gauge fixing and the Faddeev-Popov-ghost action, respectively. 
Both latter contributions are determined by
the gauge condition $F_\mu$. The gauge fixing action reads
\begin{align}
	S_\text{gf}=\01{32\pi\alpha}\int\mathrm d^4x
  \sqrt{\bar g}\,\bar g^{\mu\nu}F_\mu F_\nu\,.
\end{align}
In this work, we apply a De-Donder-type linear gauge given by
\begin{align}\label{eq:Fmu}
  F_\mu = \bar \nabla^\nu h_{\mu\nu} - \0{1+\beta}4
  \bar \nabla_\mu {h^\nu}_\nu \, ,
\end{align}
with $\beta=1$. Furthermore, we apply the 
Landau-limit of vanishing gauge parameter, $\alpha\to0$. 
The Faddeev-Popov operator corresponding to \eqref{eq:Fmu} is of the form
\begin{align}
  \mathcal{M}_{\mu\nu} =
  \bar\nabla^\rho\left(g_{\mu\nu}\nabla_\rho+g_{\rho\nu}\nabla_\mu
  \right)-\bar\nabla_\mu\nabla_\nu\,.
\end{align}
The Landau-limit $\alpha\to0$ is particularly
convenient since it provides a sharp implementation of the gauge
fixing. This assures furthermore, that the corresponding gauge-fixing parameter is
at a fixed point of the renormalisation group flow
\cite{Litim:1998qi}.

The vertex flows discussed here carry additional spacetime and
momentum indices. In order to obtain scalar flow equations for the
couplings the appropriate projection of the flows is a crucial part of
the present truncation and goes along the same lines as in
\cite{Christiansen:2015rva}. It can be summed up in a three step
procedure: 

\begin{enumerate}[(i)]
\setlength{\itemsep}{10pt}
\item We decompose $\mathcal{T}^{(n_h)}$, where $n_h$ is the number of
  variations with respect to $h$, into its momentum dependent and
  momentum independent part according to
\begin{align}\label{eq:dec_T}
  \hspace{1cm} \mathcal{T}^{(n_h)}(\mathbf{p};\Lambda_{n_h}) =
  \mathcal{T}^{(n_h)}(\mathbf{p};0)+\Lambda_{n_h}\mathcal{T}^{(n_h)}(0;1)\,.
\end{align}
In \eqref{eq:dec_T}, the first term on the right-hand side is
quadratic in the external graviton momenta $\mathbf{p}$ for the
current truncation. The second term is momentum independent.

\item From \eqref{eq:dec_T} we take the dimensionless tensors
  $\mathcal{T}^{(n_h)}(\mathbf{p};0)/\mathbf{p}^2$ and
  $\mathcal{T}^{(n_h)}(0;1)$ and separately
  multiply all spacetime-index pairs of both tensors with
  transverse-traceless projection operators $\Pi_{\TT}$. This leaves us with the
  two tensors $\mathcal{T}^{(n_h)}_\TT(\mathbf{p};0)/\mathbf{p}^2$ and
  $\mathcal{T}^{(n_h)}_\TT(0;1)$, each of them carries $2 n_h$
  spacetime indices.

\item We contract the left and the right hand side of the vertex flow
  with these two tensors, in order to obtain Lorentz-scalar
  expressions. Hereby, the tensors $\mathcal{T}^{(n_h)}_\TT
  (\mathbf{p};0)/\mathbf{p}^2$ and $\mathcal{T}^{(n_h)}_\TT (0;1)$ are
  used to project the tensorial flow onto the scalar flows of
  $G_{n_h}$ and $\Lambda_{n_h}$, respectively.
\end{enumerate}
The projection operators are detailed in Appendix \ref{app:three-point-function}. 
In addition to the spacetime indices, the vertex flows carry spinor,
flavour and colour indices. These however, can be trivially traced out
after multiplying appropriately with $\gamma$ and
$\mathbb{1}$-matrices.

After having traced out all discrete indices the resulting flow still
depends on the external field momenta $\mathbf{p}$. This dependence is
dealt with by choosing a specific kinematic configuration. Since all
vertices obey momentum conservation this choice is only relevant for
$n$-point vertices with $n\geq3$. In this work, the flow of the graviton
three-point function is the highest order vertex flow and thus it is the
only flow that needs a fixed kinematic configuration. For
the latter, we choose the maximally symmetric configuration, to wit
\begin{align} \label{eq:mom-config}
 |p_1|=|p_2|=:p\,,\qquad \vartheta=2\pi/3\,,
\end{align}
where $\vartheta$ is the angle between $p_1$ and $p_2$. Note, that
$p_3$ was eliminated using momentum conservation.  This way, both
sides of the flow equations for all vertices only depend on the scalar
momentum parameter $p$.  Note, that due to the vertex construction
\eqref{eq:vert} and the choice of regulators $R_k^\phi$ to be
specified below there are no single wavefunction renormalisations
$Z_{\phi_i}$ in the flow. Instead, the latter always enter in terms of
the corresponding anomalous dimensions $\eta_{\phi_i}$ defined by
\begin{align} \label{eq:def-anom-dim}
	\eta_{\phi_i}(p^2) := -\partial_t \ln Z_{\phi_i}(p^2)\,.
\end{align}
Consequently, the flow of a generic $\phi^n$-vertex reads
schematically
\begin{align}\label{eq:flowint}
  \text{Flow}^{(\phi^n)} = \int_q \left(\dot{r}_{\phi_i}(q^2) -
    \eta_{\phi_i}(q^2)r_{\phi_i}(q^2) \right)
  F^{(\phi^n)}_i(p,q,\dots)\,,
\end{align}
where we have defined $\text{Flow}^{(\phi^n)}$ as
\begin{align}
  \text{Flow}^{(\phi^n)}(p^2):=\0{\dot
    \Gamma^{(\phi_1\dots\phi_n)}(p^2)}{\prod_{i=1}^n \sqrt{Z_{\phi_i}(p^2)}}\,.
\end{align}
In \eqref{eq:flowint}, $r_{\phi_i}$ denotes the regulator shape
function corresponding to the field $\phi_i$ and the functions
$F^{(\phi^n)}_i$ encode the contributions of the field
$\phi_i$ to the flow of the $\phi^n$-vertex. The functions $F_i$ depend
on the external and loop momenta, $p$ and $q$, respectively, as well
as on the couplings $G$ and $\Lambda_n$. The remaining $p$-dependence
in \eqref{eq:flowint} is projected out differently, depending on the
quantity to be extracted. The momentum projection will be discussed below.

Summarising the present truncation, we consider the renormalisation
group flow for the $n$-point correlation functions in a system of
minimally-coupled gravity and matter.  To this end, we employ a vertex
expansion of the scale dependent effective action about a flat metric
background to derive flow equations for the $n$-point correlators up
to order three.  The \RG-invariant vertex dressing \eqref{eq:vert}
allows to derive independently the flows of the momentum-independent
couplings $G$, $\Lambda_2$ and $\Lambda_3$ as well as the
momentum-dependent anomalous dimensions $\eta_h(p^2)$, $\eta_c(p^2)$,
$\eta_\psi(p^2)$ and $\eta_\varphi(p^2)$. The couplings $G$ and
$\Lambda_3$ are computed from the transverse-traceless part of the
graviton three-point function in the symmetric momentum configuration.
Diffeomorphism invariant background couplings are computed from the
solution of the dynamical couplings.  Altogether, the present
truncation yields the flow of the scale dependent parameters,
\begin{align}
  \{\bar G, \bar \Lambda,
  G,\Lambda_2,\Lambda_3,\eta_h(p^2),\eta_c(p^2),\eta_\psi(p^2),\eta_\varphi(p^2)\}\,.
\end{align}

\section{Flows of correlation functions}
The properties of the given theory are completely determined by the
flows of the respective correlation functions. Thus, the latter parametrise the non-trivial
interplay between gravity and matter. Matter is known to have a
significant impact on the \UV-behaviour of quantum gravity. On the
other hand, graviton fluctuations can lead to strong correlations among
matter fields. The resulting mutual dependencies play a crucial role
for the flow of the complete system and are discussed separately in
the following sections.

The computation of correlation functions described in this section
involves the contraction of very large tensor structures. These
contractions are computed with self-developed pattern-matching
scripts.  In this context we make use of the symbolic manipulation
system {\small {\it FORM}} \cite{Kuipers:2012rf,Vermaseren:2000nd}.
\begin{figure}[t]
\centering
\includegraphics{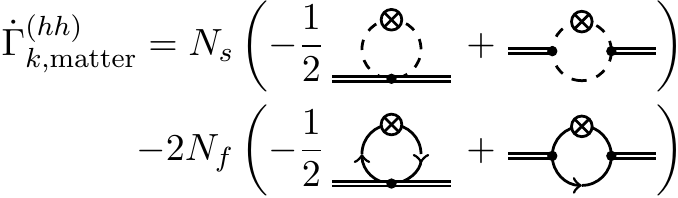}
\caption{Diagrammatic representation of the matter induced flow of the
  graviton two-point function. Double, single and dashed lines represent
  graviton, fermion and scalar propagators respectively, filled
  circles denote dressed vertices. Crossed circles are regulator
  insertions.}
\label{fig:flow_two_point}
\end{figure}
\subsection{Matter contributions to gravity flows} \label{sec:gravity-flows}
For the present analysis of quantum gravity, the gravity flows are
extracted from the dynamical graviton two-point and three-point
functions. The impact of matter manifests itself by matter loops in
the diagrammatic representation of the
flow. \autoref{fig:flow_two_point} depicts these contributions for the
flow of the graviton two-point function.  The trace over the colour
and flavour indices leads to weight factors of $N_s$ and $N_f$ for
scalar and fermion loops, respectively. The matter contributions to
$\text{Flow}^{(hh)}$ are thus proportional to $N_s$ or $N_f$. From
$\text{Flow}^{(hh)}$ we extract the flow of the graviton mass
parameter defined as $M^2:=-2\Lambda_2$ and the graviton anomalous
dimension $\eta_h$. This procedure is discussed in more detail in the
following. A complete discussion can be found in
\cite{Christiansen:2014raa}.

From \eqref{eq:vert} we obtain an equation for the
transverse-traceless graviton two-point function by contracting all
external graviton legs with $\Pi_{\TT}$. This leads to
\begin{align}\label{eq:2h}
 \Gamma^{(hh)}_{\TT}(p^2) =\01{32\pi} Z_h(p^2) (p^2+M^2)\,.
\end{align}
Taking a derivative with respect to renormalisation time $t$ and
dividing by $Z_h(p^2)$ yields
\begin{align}\label{eq:2h_flow}
  \text{Flow}_{\TT}^{(hh)}(p^2) = \01{32\pi} \left(\partial_t {M^2}
    -\eta_h(p^2)(p^2+M^2)\right)\,.
\end{align}
The right hand side of the flow equation provides an expression for
$\text{Flow}_{\TT}^{(hh)}(p^2)$, which depends solely on the couplings
and the anomalous dimensions. The resulting equation is
evaluated at two different momentum scales $p^2$. Subtracting these
two equations from each other allows for an unambiguous extraction of
$\partial_t {M^2}$ and $\eta_h(p^2)$. We call this procedure bilocal
momentum projection, it is applied for gravity in
\cite{Christiansen:2012rx,Christiansen:2014raa,Christiansen:2015rva}. 

We extract the ghost anomalous dimension $\eta_c(p^2)$ from the transverse
part of the ghost two-point function. The significance of the ghost
contributions and details on their extraction are explained in
\cite{Christiansen:2012rx,Christiansen:2014raa} and the explicit form
is given in Appendix \ref{app:anom_dim}.

\begin{figure}[t]
\includegraphics[width=8.5cm]{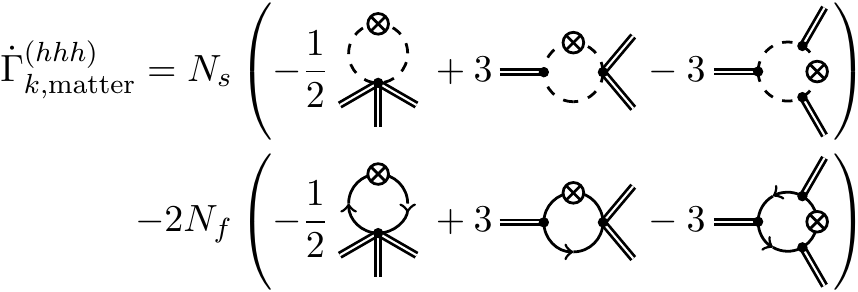}
\caption{Diagrammatic representation of the matter induced flow of the
  three-graviton vertex. Double, single and dashed lines represent
  graviton, fermion and scalar propagators respectively, filled
  circles denote dressed vertices. Crossed circles are regulator
  insertions. All diagrams are symmetrised with respect to the
  interchange of external momenta $\mathbf{p}$.}
\label{fig:flow_three_point}
\end{figure}

The matter contributions to the flow for the graviton three-point
function parametrise the impact of matter on the dynamical
gravitational couplings $g$ and
$\lambda_3$. \autoref{fig:flow_three_point} shows the matter
contributions arising via loops in the diagrammatic representation.
Again, the multiplicity of the matter-loops leads to contributions to
$\text{Flow}^{(hhh)}$ proportional to $N_s$ and $N_f$. The flow for
$G$ and $\Lambda_3$ is extracted in a vein, similar to the extraction of
$\eta_h$ and $\partial_t {M^2}$ from the graviton two-point
function. Projecting the flow of the three-graviton vertex on the
transverse-traceless contributions of the classical tensor structures
as described above and evaluating the kinematic configuration at the
symmetric point as described in \eqref{eq:mom-config}, yields
equations of the type
\begin{align} \label{eq:Gamma_3h} \Gamma^{(hhh)}_{\TT,i}=
  Z_h^{3/2}(p^2)\, G^{1/2}
  \left(\mathcal{N}_i\,p^2+\mathcal{M}_i\,\Lambda_3 \right)\,.
\end{align}
with $i = G,\Lambda$, for the projection on the tensor structures of
$G$ and $\Lambda_3$, respectively. The factors $\mathcal{N}_i$ and
$\mathcal{M}_i$ arise from the tensor projection. They depend on the
kinematic configuration and are given explicitly in Appendix
\ref{app:three-point-function} for the symmetric momentum
configuration \eqref{eq:mom-config}. Taking a scale derivative and
rearranging leads to
\begin{multline}\label{eq:3h_flow}
  \0{2}{\sqrt{G}}\,\text{Flow}^{(hhh)}_{\TT,i}=2\mathcal{M}_i\,\partial_t \Lambda_3  \\
  - \left[\eta_G+3 \eta_h(p^2)\right]
  \left(\mathcal{N}_i\,p^2+\mathcal{M}_i\,\Lambda_3 \right)\,, 
\end{multline}
with $\eta_G=-\partial_t \ln G$. Note, that \eqref{eq:3h_flow} is
structurally very similar to \eqref{eq:2h_flow}. For the extraction of
the flows for the couplings $G$ and $\Lambda_3$ we apply the bilocal
momentum projection discussed before. Thus, we evaluate the flow of
$G$ at $p=k$ as well as at $p=0$ and subtract both equations from each
other. Since the term proportional to $\partial_t \Lambda_3$ in
\eqref{eq:3h_flow} is momentum independent, it drops out upon the
subtraction thus leaving us an equation for $\partial_t G$. For the
flow of $\Lambda_3$ it is then sufficient to evaluate
\eqref{eq:3h_flow} (with $i=\Lambda$) at vanishing external momentum
$p=0$. The resulting flow equations are identical to the ones in
\cite{Christiansen:2015rva} and are given in Appendix
\ref{app:three-point-function}.

\subsection{Gravity contributions to matter flows}
In the matter sector, we consider the flows of the matter two-point
functions. Since we do not admit matter self-interactions within the
given truncation, these flows are driven solely by gravity-matter
interactions. Furthermore, the matter fields are treated as massless,
which is a good approximation for studies of the \UV-behaviour of the
theory. Consequently, the only quantities that are extracted here, are
the matter anomalous dimensions. The effective action constructed from
\eqref{eq:EH} is diagonal in both the colour and the flavour indices,
$i$ and $k$, respectively. We treat all scalars and all fermions on 
equal footing, providing them with one anomalous dimension
for each of the field species, $\eta_\varphi(p^2)$ and $\eta_\psi(p^2)$,
respectively. This allows for an extraction of the matter anomalous
dimension from one representative field, since
$\text{Flow}^{(\varphi_k\varphi_l)}=\delta_{kl}\text{Flow}^{(\varphi_k\varphi_k)}$
and $\text{Flow}^{(\bar \psi_i\psi_j)}=\delta_{ij}\text{Flow}^{(\bar
  \psi_i\psi_i)}$. Consequently, we drop the colour and flavour
indices in the flows of the scalar and fermion two-point functions.

\begin{figure}[t]
\centering
\includegraphics{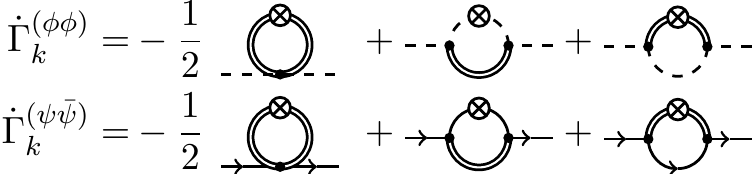}
\caption{Diagrammatic representation of the gravitationally induced
  flows of the matter-two-point-functions.  Double, single and dashed
  lines represent graviton, fermion and scalar propagators
  respectively, filled circles denote dressed vertices. Crossed
  circles are regulator insertions.}
\label{fig:flow_matter}
\end{figure}

\autoref{fig:flow_matter} depicts the flows of the matter two-point
functions in diagrammatic representation which constitute the
respective right hand sides of flow equation.
From these flows we extract the matter anomalous dimensions. For the
scalar fields the left hand side is given by
\begin{align}\label{eq:2phi}
 \text{Flow}^{(\varphi\varphi)}(p^2)=-p^2\eta_\varphi(p^2)\,,
\end{align}
in complete analogy to the equation for the transverse-traceless
graviton two-point function, \eqref{eq:2h}. For the fermions we have
the additional spinor structure which needs to be eliminated in order
to obtain a Lorentz-scalar expression. The flow for the fermion two-point function reads
\begin{align}
 \text{Flow}^{(\bar \psi\psi)}(p^2)=-i\slashed p\, \eta_\psi(p^2)\,.
\end{align}
By multiplying this expression with $\slashed p$ and taking the trace
over the spinor indices we obtain an expression, which is identical to
\eqref{eq:2h} and \eqref{eq:2phi} up to prefactors, to wit
\begin{align}\label{eq:2psi}
  \text{tr}(\slashed p\,\text{Flow}^{(\bar
    \psi\psi)})(p^2)=-d\, i\, p^2\eta_\psi(p^2)\,.
\end{align}
Here $d$ is the dimension of spinor space, which we set to $d=4$
throughout. Since \eqref{eq:2h}, \eqref{eq:2phi} and \eqref{eq:2psi}
are of the same form, we apply the same bilocal momentum projection
for the extraction of the respective momentum dependent anomalous
dimensions. This crucial procedure is discussed in more detail in the
next section.

\subsection{Anomalous Dimensions} \label{sec:anom-dim}
Each of the field species is equipped with an anomalous dimension $\eta_{\phi_i}(p^2)$. 
The latter are extracted from the flow of the respective field's two point function.
In the context of heat-kernel methods, the anomalous dimensions are often referred to as
`\RG{} improvement' \cite{Groh:2010ta,Eichhorn:2010tb,Codello:2013fpa,Dona:2013qba}.
In this work, they arise naturally from the truncation and as further improvement we 
keep an approximated momentum dependence of the anomalous dimension, similar 
to \cite{Christiansen:2014raa,Christiansen:2015rva}.

The expressions
\eqref{eq:flowint}, \eqref{eq:2phi} and \eqref{eq:2psi}, together with
the bilocal momentum projection lead to a coupled system
of Fredholm integral equations for the anomalous dimensions
$\vec\eta_\phi=(\eta_h,\eta_c,\eta_\psi,\eta_\varphi)$. 
The specific form of the latter is
given in Appendix \ref{app:anom_dim}. It can be written as
\begin{align}\label{eq:inteq_gen}
  \vec \eta_\phi(p^2) = \vec A(p^2,G,M^2,\Lambda_3)+ \vec
  B(p^2,G,M^2,\Lambda_3)[\vec \eta_\phi]\,,
\end{align}
where $\vec A$ and $\vec B$ are momentum-integral expressions. As the
square brackets suggest, $\vec B$ is a functional of $\vec
\eta_\phi(q^2)$. Equation \eqref{eq:inteq_gen} can be solved
iteratively which is, however, computationally very expensive since it
is a coupled system of four equations. In order to get a handle on the
solution of \eqref{eq:inteq_gen}, we evaluate the anomalous dimension
in $\vec B$ at $k^2$ and move $\eta_\phi(k^2)$ in front of the
integrals. This is a good approximation because all integrals of the
type \eqref{eq:flowint} are sharply peaked around $q=k$. This feature arises
due to the factor of $q^3$ from the integral measure in $d=4$ spherical
coordinates. Since $\vec B$ is linear in $\vec\eta_\phi$, we can now
write it as a matrix $C$ multiplying the vector $\vec
\eta_\phi(k^2)$. Hence, \eqref{eq:inteq_gen} simplifies to
\begin{align}\label{eq:inteq_lin}
  \vec \eta_\phi(p^2) \approx \vec A(p^2,G,M^2,\Lambda_3)+
  C(p^2,G,M^2,\Lambda_3)\,\vec \eta_\phi(k^2)\,.
\end{align}
We now evaluate the latter equation at $p=k$ in order to obtain an
expression for $\vec\eta_\phi(k^2)$. The result $\vec \eta_\phi(k^2)$ is
substituted back into the momentum-dependent equation
\eqref{eq:inteq_lin}.  This way, we obtain anomalous dimensions with
an approximated momentum dependence. Note, that the latter
approximation is considerably better than the assumption of
momentum-independent anomalous dimensions, since we evaluate the
functional dependence on $\vec \eta_\phi$ at the peak position of the
integrals. In particular, this procedure allows for a distinction of
$\vec \eta_\phi(k^2)$ and $\vec \eta_\phi(0)$, which is important
since they both appear explicitly in the flow equations \eqref{eq:analyticflows},
due to the bilocal momentum projection. We show in
\autoref{sec:gravity-results} that our approximation is
justified for the case without matter via comparison with the results
from \cite{Christiansen:2015rva}.

As an interesting fact, the scalar anomalous dimension
$\eta_\varphi(p^2)$ vanishes for the given graviton gauge. Generally,
the scalar anomalous dimension comprises a term which is proportional
to the scalar mass and one mass-independent term. The latter vanishes
for the used harmonic gauge. Obviously, the former term vanishes for
massless scalars which we consider here, leaving us with a vanishing
scalar anomalous dimension $\eta_\varphi(p^2)=0$. Note that this is
only the case for the scalar anomalous dimension in this particular
gauge, for all other gauges $\eta_\varphi(p^2)$ is not equal to zero.

\subsection{Anomalous dimensions and bounds for the
    generic class of regulators}\label{sec:bounds-anom-dim}
As part of the truncation, we choose a generic class
  of regulators $R^\phi_k$, that are proportional to the
corresponding two-point function, i.e.
\begin{align}\label{eq:reg_def}
  R^\phi_k(p^2)=\Gamma^{(\phi\phi)}_k(p^2) r^\phi_k(p^2)\bigg|_{M^2=0}\,,
\end{align}
in momentum space, where $r^\phi_k(p^2)$ is the regulator shape
function. Note, that the evaluation of the two-point function at $M^2=0$ ensures
that only the momentum dependent part of the latter enters for the class of
regulators defined by \eqref{eq:reg_def}. Since the effective graviton mass
$M^2$ is the only mass parameter in the present truncation the above definition implies that
$\Gamma^{(\phi\phi)}_k(p^2)|_{M^2=0}$ is either the full
two-point function $\Gamma^{(\phi\phi)}_k(p^2)$, or, in case of the graviton field, its
momentum-dependent part, i.e.\ $\Gamma^{(hh)}_{k,\TT}(p^2)|_{M^2=0}=(32\pi)^{-1} Z_h(p^2)p^2$, see
\eq{eq:2h}. This generic class covers the regulator choices in the
literature, and implements the correct renormalisation group scaling
of the effective action as discussed in
\cite{Pawlowski:2001df,Pawlowski:2005xe,Christiansen:2014raa}. It
provides a RG-covariant infrared regularisation of the spectral values
of the two-point function, and is hence called RG- or spectrally
adjusted, \cite{Pawlowski:2001df,Gies:2002af,Pawlowski:2005xe}.  It
implies in particular, that the regulator is proportional to the
corresponding field's wavefunction renormalisation via the dependence
of $R^\phi_k$ on the two-point function. Thus, the present choice
leads to closed equations in terms of the anomalous
dimensions. However, for large $\eta_\phi$ the choice
\eqref{eq:reg_def} leads to a peculiar \RG-scaling of $R^\phi_k$ in
the \UV. From the path integral point of view one expects a \UV{}
scaling with
\begin{align}\label{eq:wrong_scal}
 \lim_{k\to\infty}R^\phi_k(p)\sim \lim_{k\to\infty}Z_\phi k^i \ra \infty\,,
\end{align}
for all momenta $p$. In \eqref{eq:wrong_scal} we have $i=1$ for fermions
and $i=2$ for all other fields. Equation \eqref{eq:wrong_scal} entails that the
regulator diverges in the \UV, and the related momentum modes in the
path integral are suppressed.  Since the wavefunction renormalisation
behaves like $Z_\phi\sim k^{-\eta_\phi}$ for large $k$, equation
\eqref{eq:wrong_scal} is violated if the anomalous dimensions exceed
the constraints
\begin{align}\label{eq:anomdim_constr}
  \eta_h<2 \,,\quad\quad \eta_c<2 \,,\quad\quad \eta_\varphi<2
  \,,\quad\quad \eta_\psi<1\,.
\end{align}
Hence, if one of the bounds in \eqref{eq:anomdim_constr} is violated,
the respective regulator vanishes in the \UV. In the spirit of the
above path integral picture this may imply a decrease of the effective
cutoff scale for the respective field, and hence a flow towards the
\IR. Note however, that this is far from being clear from the flow
equation itself. For example, with the regulator \eqref{eq:reg_def}
the \TText-component of the graviton propagator is proportional to
\begin{align}\label{eq:propspec} 
  \0{1}{Z_{h}(p^2)}\0{1}{(p^2(1+r_h)+M^2)}\,,
\end{align}
which implies a spectral, RG-covariant regularisation of the momentum
modes of the full propagator, as discussed above. We conclude that if
the bounds in \eqref{eq:anomdim_constr} are exceeded, the regulator
may not suppress field modes in the \UV{} properly. Indeed, if the
anomalous dimension are large enough, this does not only lead to a
decreasing regulator, but also $\partial_t R^\phi_k$ turns
negative. This can be seen from the schematic expression
\begin{align} \label{eq:t-der-reg}
  \partial_t R_k^\phi\sim Z_\phi(\dot r_k^\phi(p^2)-\eta_\phi
  r^\phi_k(p^2))\,.
\end{align}
The second term in equation \eqref{eq:t-der-reg} exceeds the first one
for $p/k\to0$ exactly at the critical values given in
\eqref{eq:anomdim_constr}. Still, this is not sufficient to change the
sign of the respective diagrams, which involves an integration over
all momenta. However, for an even larger anomalous dimensions,
$\eta_{\text{\tiny{sign}}}>2$, the sign of the respective diagrams
changes. In the path integral interpretation introduced above this
change of sign signals the global change from a \UV-flow to a \IR-
flow for the respective diagram. Naturally, this bound depends on the
shape function of the regulator. For the present approximation, the
first diagrams switch sign at $\eta_{\text{\tiny{sign}}}=4$. This is
already visible in the analytic, reduced, approximation derived later,
see \eqref{eq:analyticflows}. Note also, that the sign of diagrams
does not change for $Z_h$-independent regulators. Accordingly, for
$\eta_h>\eta_{\text{\tiny{sign}}}$ we have a regulator-dependence of
the sign of diagrams, which has a qualitative impact on the physics
under discussion. Hence, for $\eta_h>\eta_{\text{\tiny{sign}}}$ the
present approximation breaks down completely. In the present work,
however, we resort to the stricter, shape-function-independent bound
\eq{eq:anomdim_constr}.

In summary, it is clear that if the bounds
\eqref{eq:anomdim_constr} are violated, additional investigations
of the regulator-dependence, and hence of the reliability of the
present approximation are required. Note however, that small anomalous
dimensions, that obey \eqref{eq:anomdim_constr}, do by no means
guarantee the convergence of the results with respect to an extension
of the truncation. Such a convergence study requires the inclusion of
higher order operators and detailed regulator studies and is deferred
to future work.

\section{Results}\label{sec:results}
In this section, the results of the above presented setups are
displayed. As a main result, within the validity
  bounds for the chosen generic class of regulators, we do not find
an upper limit for the numbers of scalars and fermions that are
compatible with the asymptotic safety scenario.

For the analysis we employ regulators of the type given in
\eqref{eq:reg_def} and use a Litim-type shape function
\cite{Litim:2000ci} that is, $\sqrt{x} r(x) =(1-\sqrt{x})\theta(1-x)$
for fermions and $x r(x) =(1-x)\theta(1-x)$ for all other fields. We
close the flow equations with the identification
$\Lambda_5=\Lambda_4=\Lambda_3$. Furthermore, we work with the
dimensionless quantities
\begin{align} \label{eq:dimless} &g:= G k^2\,, \;\;\;\;\;\;\;
  \mu:=M^2k^{-2}\,, \;\;\;\;\;\;\; \lambda_3:=\Lambda_3 k^{-2}\,.
\end{align}

\subsection{Pure gravity} \label{sec:gravity-results}
In order to study the \UV{} behaviour of quantum gravity
interacting with matter, we start from the \UV{} fixed point of pure
quantum gravity found in \cite{Christiansen:2015rva} and study the
deformation of this particular fixed point by the matter
content. To that end, we rederive the results for the pure gravity case with
the approximated momentum dependence of the anomalous dimensions
discussed in \autoref{sec:anom-dim}. We compare these findings with the results
in \cite{Christiansen:2015rva}, where the full momentum dependence of
the latter was considered. The fixed point values for the pure-gravity
system in the present approximation read
\begin{subequations}\label{eq:FPparam}
\begin{align}\label{eq:uvfp}
 (g^*,\mu^*,\lambda_3^*)= (0.62,-0.57,0.095)\,,
\end{align}
with the critical exponents $\theta_1$, $\theta_2$ and $\theta_3$ given by
\begin{align}\label{eq:critexp}
 (\theta_{1/2},\theta_3)=(-1.3\pm4.1\, i ,12)\,.
\end{align}
\end{subequations}
These fixed point values are in agreement with
\cite{Christiansen:2015rva} within an error of $6\%$
($15\%$ for the critical exponents). This justifies the approximations
described in \autoref{sec:anom-dim}. The deformation of the fixed
point \eqref{eq:FPparam} is calculated while successively increasing
the number of scalars and fermions, $N_s$ and $N_f$,
respectively. This way, we analytically continue the fixed point of
the pure gravity system towards a theory of quantum gravity and
matter, which contains $N_s$ scalars and $N_f$ fermions. Although
$N_f$ and $N_s$ are (half-)integers in the physical sense, we treat
them as continuous deformation parameters for this analysis. With this
procedure we simulate the generic effect of gravity-matter
interactions on gravity-theories. First, we analyse the influence of
scalars and fermions separately before we briefly discuss combined
system of both matter types.

\subsection{Scalars}\label{sec:scalars}
\begin{figure*}[t]
\includegraphics[width=.99\textwidth]{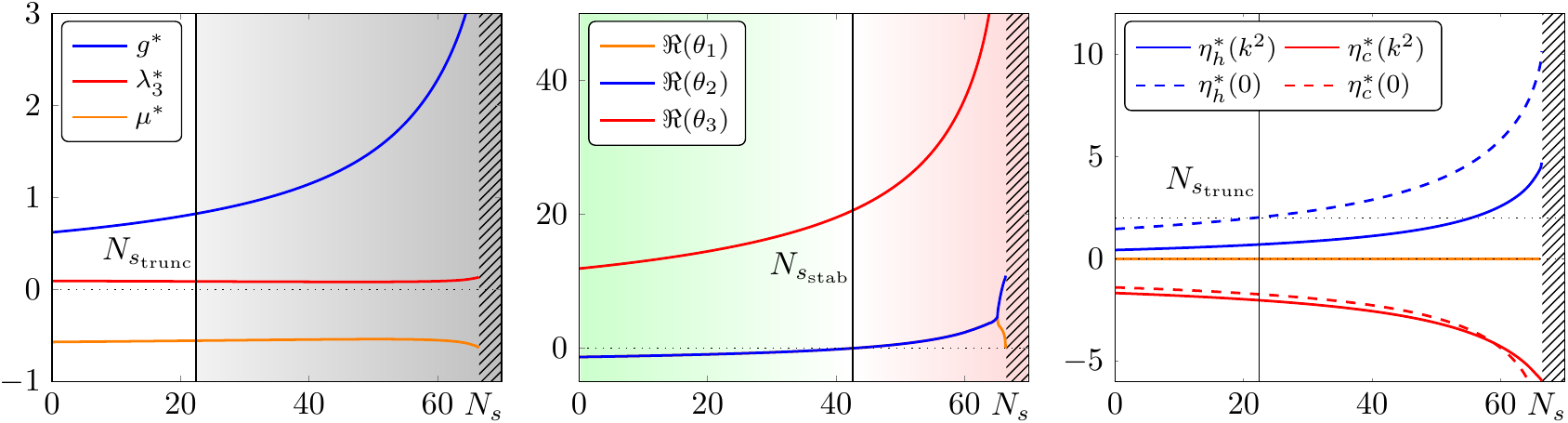}
\caption{Fixed point values (left), real parts of the critical
  exponents (middle), and the anomalous dimensions evaluated at $p=k$
  and $p=0$ (right) as functions of the number of scalars, $N_s$,
  respectively. The grey-shaded area in the left panel indicates where
  the regulator lies outside the reliability bounds defined in
  \autoref{sec:bounds-anom-dim} due to a large graviton anomalous
  dimension (see right panel). The corresponding limiting number or
  scalars is given by $N_{s_\text{trunc}}$. The hatched regions in all
  three panels correspond to the $N_s$-regime where the \UV{} fixed
  point does not exist. The green and red areas in the middle panel
  denote the region where the fixed point exhibits \UV{}-attractive
  direction and the region where it is fully repulsive,
  respectively. $N_{s_\text{stab}}$ is the corresponding critical
  number. The colours of the curves in the middle panel indicate with
  which coupling of the left panel the corresponding eigenvector has
  the largest overlap. The anomalous dimension of the scalar
  $\eta_\varphi$ (right panel) is zero due to the given graviton
  gauge.}
  \label{fig:NS_findiff_w_eta}
\end{figure*}
We first consider the case $N_f=0$, $N_s>0$, thus a theory of $N_s$
scalars minimally coupled to gravity. Note again, that in the present
approach, we neglect the influence of scalar self-interactions in the
action \eqref{eq:EH}. Detailed analyses of the potential impact of
matter-matter couplings can be found in
e.g. \cite{Eichhorn:2011pc,Eichhorn:2012va, Henz:2013oxa}.

Before analysing the full numerical flow equations we try to
anticipate the result from the analytic flow equations
\eqref{eq:analyticflows} without anomalous dimensions. For $N_f=0$,
$N_s>0$ and $\vec \eta_\phi=0$ the latter equations read
\begin{align} \label{eq:ns_flows}
  \dot g =& +2g + \beta_{g_{\text{Gravity}}}- \0{43}{570 \pi} g^2 N_s\,, \notag\\
  \dot \mu =& -2 \mu +  \beta_{\mu_{\text{Gravity}}} + \01{12 \pi} g N_s\,, \notag\\
  \dot \lambda_3 =&-2 \lambda_3 + \beta_{\lambda_{3,\text{Gravity}}}
  -\frac{1}{60\pi}\left( 1 -\frac{43}{19}\lambda_3\right)g N_s\,.
\end{align}
In this set of equation we have split the running of the dimensionless
couplings $(g,\mu,\lambda_3)$ into the canonical running, the
contribution from graviton and ghost loops, and the contribution from
scalar loops, in this ordering. In the following, we analyse whether
the respective signs of the contributions potentially stabilise or
destabilise the \UV{} fixed point. A matter contribution to a given
flow equation potentially destabilises the \UV{} fixed point of the
pure gravity system if it has the same sign as the canonical
running. In this case, the contributions from graviton and ghost loops
need to increase in to order to compensate for the 
matter contribution and, thus, allow for a gravity-matter fixed
point. Conversely, if the canonical running and the matter
contributions have the opposite sign we consider the matter
contributions to potentially stabilise the fixed point. Further, we
argue that the matter contribution to the running of $\mu$ has the
largest impact on the flow compared to the other equations of the
system \eqref{eq:ns_flows}.

Using the above notion, the scalar contribution to $\partial_t g$
potentially stabilises the fixed point, since the canonical running of
$g$ is positive and the $N_s$-dependent term has a negative sign.  The
positive sign of the $N_s$-term in $\partial_t \mu$ potentially
destabilises the fixed point, since we have found $\mu^*<0$ in the
pure gravity case (see \eqref{eq:uvfp}). Moreover, the contribution to
$\partial_t \lambda_3$ is potentially destabilising, since we consider
a positive and small $\lambda_3$ as in \eqref{eq:uvfp}. The behaviour
is opposite for $\lambda_3>19/43$ and for $\lambda_3<0$.

We note that the flow equation for $\mu$ has the largest impact on the
complete system \eqref{eq:ns_flows}. For one, that is because $\mu$ is
the effective mass parameter of the graviton and, consequently,
appears in all diagrams with graviton contributions in the loops. The
second reason is that the fixed point value $\mu^*$ for the pure
gravity system is close to -1. The $\mu$-contributions to the flow
equations generally take the form $(1+\mu)^{-n}$ with
$n\geq1$. Perturbations of $\mu$ are therefore strongly amplified if
$\mu$ is close to $-1$. To see this we expand the general form of the
$\mu$-contributions around $-1/2$, namely $\mu=-1/2+\epsilon$, which
is approximately the fixed point value of the pure gravity system (see
\eqref{eq:uvfp}). The general form of the $\mu$-contributions is now
given by
\begin{align} \label{eq:mu-effect}
 \frac{1}{(1+\mu)^n}=\frac{2^n}{(1+2\epsilon)^n}\approx 2^n(1-2n\epsilon)\,,
\end{align}
which suggests that small perturbations of $\mu$ around $-1/2$ are
amplified by a factor of $2n$ compared to contributions of order one
which appear linearly in the numerators . Using a Litim-type regulator
we obtain terms of the latter type in $\partial_t g$ up to $n=5$. For
these terms perturbations of $\mu$ around $-1/2$ are amplified by 10
compared to the linear quantities of order one. The impact of $\mu$ on
the flow \eqref{eq:ns_flows} becomes even larger, the closer $\mu$ is
driven towards -1. For $\partial_t \lambda_3$ this argument is
additionally supported by the smaller scalar contribution to
$\partial_t\lambda_3$ compared to the respective contributions to
$\partial_t g$ and $\partial_t \mu$. This also compensates for the
fact that the fixed point value in the pure gravity case is
$\lambda_3^*\approx1/10$ and therefore not of order one. For these
reasons, the scalar contributions in the flow of $\mu$ have the
largest impact on the system \eqref{eq:ns_flows}.

In summary, we anticipate that the inclusion of scalar degrees of
freedom potentially destabilises the \UV{} fixed point. Hence, the
gravity contributions in \eqref{eq:ns_flows} must increase in order to
compensate the destabilising $N_s$-contributions. This suggests that
the couplings $g^*$ and $\lambda_3^*$ must increase with increasing
$N_s$.

We now turn to the discussion of the \UV{} fixed point for a varying
number of scalars $N_s$ in the full truncation. The left panel in
\autoref{fig:NS_findiff_w_eta} shows the fixed point values of the
dynamical quantities $(g^*,\mu^*,\lambda_3^*)$ of the system as a
function of $N_s$. All fixed point values are continuous functions of
the number of scalars in the regime $0 \leq N_s \leq
66.4=:N_{s_\text{max}}$. Outside this regime (hatched area), the fixed
point disappears, thus, spoiling asymptotic safety of the
corresponding theory. For $0 \leq N_s \leq 66.4=:N_{s_\text{max}}$,
the fixed point value for the gravitational coupling $g^*$ (blue
curve) increases with increasing $N_s$, as conjectured from the
analytic equations \eqref{eq:ns_flows}. Both, $\lambda_3^*$ and
$\mu^*$, depicted in red and orange, respectively, remain almost
constant, exhibiting only minor variations close to
$N_{s_\text{max}}$. The grey-shaded area in the left panel indicates
where the regulator lies outside the reliability
bounds defined in \autoref{sec:bounds-anom-dim} due to a large
graviton anomalous dimension (see right panel). The corresponding
limiting number or scalars is given by $N_{s_\text{trunc}}$.

\begin{figure*}[tp!]
\centering
 \includegraphics[width=.99\textwidth]{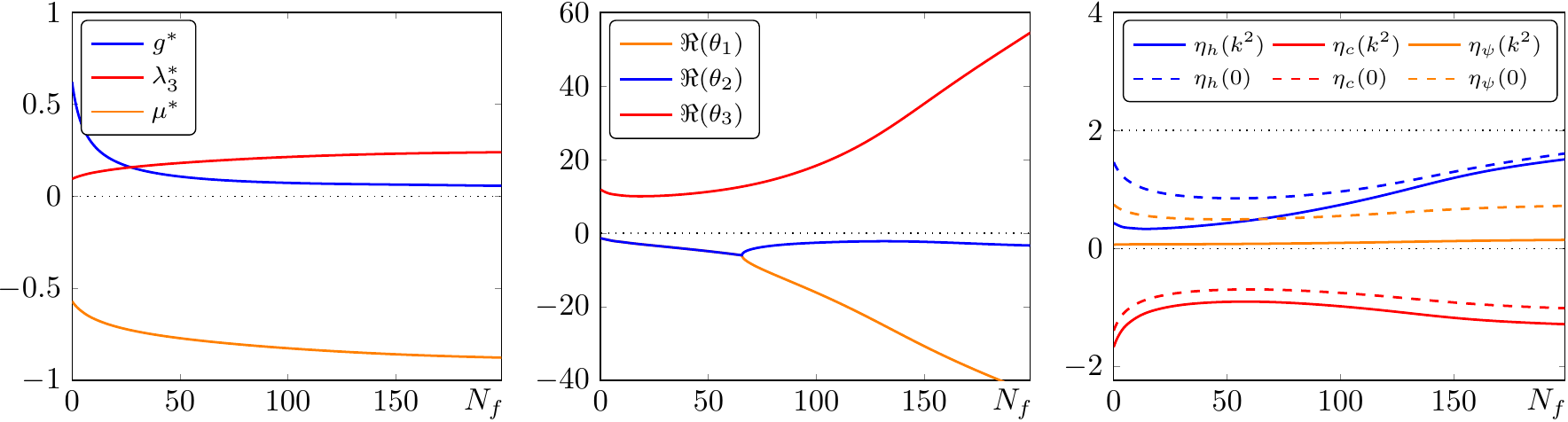}
 \caption{Fixed point values (left), real parts of the critical
   exponents (middle), and the anomalous dimensions evaluated at $p=k$
   and $p=0$ (right) as functions of the number of fermions, $N_f$,
   respectively. The colours of the curves in the middle panel
   indicate with which coupling of the left panel the corresponding
   eigenvector has the largest overlap. All quantities remain
   well-behaved for any number of fermions. In particular, the fixed
   point stays attractive (middle panel) and the anomalous dimensions
   remain small (right panel).}
  \label{fig:NF_findiff_w_eta}
\end{figure*}

The middle panel in \autoref{fig:NS_findiff_w_eta} depicts the real
parts of the critical exponents of the fixed point
$(\Re(\theta_1),\Re(\theta_2),\Re(\theta_3))$ as functions of
$N_s$. The colours of the curves are chosen such that the
corresponding eigenvectors have the largest overlap with the coupling
of the same colour in the left panel. All critical exponents increase
with increasing $N_s$. The real part of the complex conjugate pair of
eigenvalues $\Re(\theta_{1,2})$, represented by the blue and orange
curves, changes sign at $N_{s_\text{stab}}=42.6$. Consequently, the
green and red areas correspond to $N_s$-regimes where the fixed point
exhibits attractive directions and regimes where it is fully \UV{}
repulsive, respectively. In the regime
$N_{s_\text{stab}}<N_s<N_{s_\text{max}}$ the fixed point is fully
\UV-repulsive (red area). Furthermore, we observe that $\theta_3$
takes large values for large $N_s$, which we see as further evidence
for the insufficiency of the truncation in this regime
\cite{Falls:2013bv, Falls:2014tra}.

The right panel in \autoref{fig:NS_findiff_w_eta} shows the anomalous
dimensions of all involved fields evaluated at the fixed point and at
the peak of the loop integrals, $p=k$, as well as at vanishing
momentum, $p=0$. As discussed in \autoref{sec:anom-dim}, the scalar
anomalous dimension $\eta_\varphi(p^2)$ (orange curve) is zero for all
$p$ within the chosen gravity-gauge. In consequence, it does not
appear explicitly in the legend of the panel. The graviton anomalous
dimensions $\eta_h(k^2)$ and $\eta_h(0)$ both increase with increasing
$N_s$ due to the increase of $g^*$. At $N_{s_\text{trunc}}=21.5$,
$\eta_h(0)$ exceeds the critical value of $\eta_{h_\text{crit}}=2$,
discussed in \autoref{sec:anom-dim}. Consequently, in the regime
$N_{s_\text{trunc}}\leq N_s \leq N_{s_\text{max}}$, the graviton
anomalous dimension has exceeded the reliability bounds of the generic
regulator class used here and we lose control over the suppression of
graviton field modes by the regulator.

In summary, we draw the conclusion that within our truncation the
inclusion of up to $N_s\approx21$ scalars is consistent with the
asymptotic safety scenario of quantum gravity.  We also find that
beyond this limit, our truncation exhibits a large graviton anomalous
dimension beyond the critical value defined in
\eqref{eq:anomdim_constr}. This suggests that the truncation should be
improved in order to draw definite conclusions about the regime
$N_s>N_{s_\text{trunc}}$. Therefore, the limits $N_{s_\text{stab}}$
and $N_{s_\text{max}}$ found above, should be treated with caution as
they could be artefacts of the present truncation. 

The $N_s$-dependence of the couplings shown in
\autoref{fig:NS_findiff_w_eta} is qualitatively different from that in
\cite{Percacci:2002ie,Dona:2013qba,Dona:2014pla}. This qualitative
difference is also present in the fermion system discussed in the next
section. A detailed comparison and evaluation of the reliability of
the corresponding approximations is deferred to
\autoref{sec:background-field-approx}.

We conclude this analysis with a brief discussion of the bound
$\eta_{\text{\tiny{sign}}}$. We have argued in
\autoref{sec:anom-dim} that for $\eta_h>\eta_{\text{\tiny{sign}}}$ the
present approximation breaks down completely as the sign of diagrams
is regulator-dependent. For the regulator used here, see
\autoref{sec:results}, we have $\eta_{\text{\tiny{sign}}}=4$, see also
\eqref{eq:analyticflows}. Then the approximation breaks down for
$N_{s_\text{sign}}\approx 65.4$ that is below but close to
$N_{s_\text{max}}$. As discussed in \autoref{sec:anom-dim},
$N_{s_\text{sign}}$ signals the global change from a \UV-flow to an
\IR-flow for the respective diagram and hence a mixed \UV-\IR--flow. 
Naturally, we expect the loss of the \UV{} fixed point for such a flow.

\subsection{Fermions}\label{sec:fermions}
In this section we discuss the effect of minimally coupled fermions,
thus $N_f>0$ and $N_s=0$ in our
notation. As before, matter-self interactions are neglected.

Again, we first analyse the generic behaviour of the system
of analytic flow equations (see Appendix \ref{app:analytic-eq}) with
the simplification $\vec \eta_\phi=0$. To that end, we again divide
the flow into canonical running, gravity- and ghost-loop
contributions, and matter-loop terms. Consequently, the latter
equations read
\begin{align}\label{eq:nf_flows}
  \dot g =& + 2g +\beta_{g_{\text{Gravity}}}- \0{3599}{11400\pi} g^2  N_f\,,\notag\\
  \dot \mu =& -2 \mu +\beta_{\mu_{\text{Gravity}}}-  \08{9 \pi}  g N_f\,,\\
  \dot \lambda_3 =& - 2 \lambda_3 + \beta_{\lambda_{3,\text{Gravity}}}
  + \01{20\pi}\left(\0{47}{7} + \0{3599}{1140} \lambda_3 \right) g
  N_f\,. \notag
\end{align}
Using the notion introduced in the last section, we conclude that the fermionic
contributions to $\partial_t g$ and $\partial_t \mu$ potentially stabilise
the \UV{} fixed point since they have signs opposite to the respective
canonical running. The fermionic contribution to
$\partial_t\lambda_3$, by contrast, is potentially
destabilising. As we argued in the last section, the matter
contribution to $\partial_t \mu$ is the most relevant one. Therefore,
we expect that the fermion-gravity system remains stable under the
increase of $N_f$. In particular, we expect smaller values for $g^*$
for increasing $N_f$.

We turn now to the full numerical equations with momentum dependent
anomalous dimensions.  The left panel in
\autoref{fig:NF_findiff_w_eta} shows the fixed point values of the
dynamical quantities $(g^*,\mu^*,\lambda_3^*)$ as functions of the
number of fermions $N_f$.  The fixed point value of $g$ decreases with
increasing $N_f$ and approaches $g^*\to0$ asymptotically. At the same
time, $\mu^*$ decreases with increasing $N_f$ and approaches
$\mu^*\to\mu_{\text{pole}}=-1$ for $N_f\to\infty$. The fixed point
value of $\lambda_3$ increases slightly with $N_f$ and is driven
towards an asymptotic value of $\lambda_3^*\approx1/4$. It is
important to note that the crucial negative sign of the fermionic
contribution to $\partial_t\mu$, which is the same as in the analytic
equations \eqref{eq:nf_flows}, gives rise to an interesting
stabilising effect: Since we start with a negative $\mu^*$ for $N_f=0$
the negative fermionic contribution in $\partial_t\mu$ drives $\mu^*$
towards more negative $\mu$ and therefore closer towards the
propagator pole at $\mu_{\text{pole}}=-1$. This increases the
contributions from graviton loops, which have the opposite sign
compared to the fermionic terms to $\partial_t \mu$. Thus, the latter
contributions cancel each other and the system settles at small values
of $g^*$.

The middle panel in \autoref{fig:NF_findiff_w_eta} depicts the real
parts of the critical exponents of the fixed point
$(\Re(\theta_1),\Re(\theta_2),\Re(\theta_3))$ as functions of
$N_f$. The colours are chosen such that the corresponding eigenvectors
have the largest overlap with the coupling of the same colour in the
left panel. The critical exponent of the repulsive direction
$\theta_3$ first decreases slightly and then increases to large
values. The other two critical exponents $\theta_{1,2}$ form a complex
conjugate pair with a decreasing real part until they reach
$N_f=65.5$. For $N_f>65.5$ all critical exponents are real. In this
regime, $\theta_1$ decreases to smaller values, while $\theta_2$
remains almost constant. The large absolute values of the critical
exponents $\theta_1$ and $\theta_3$ indicate, similar to the scalar
case, the necessity to extend the given truncation. Large critical
exponents appear in particular for large numbers of fermions. 

The right panel in \autoref{fig:NF_findiff_w_eta} shows the anomalous
dimensions of the graviton, the ghost, and the fermion, $\eta_h$,
$\eta_c$ and $\eta_\psi$, respectively, evaluated at the fixed
point. The anomalous dimensions are evaluated at the relevant momentum
scales, $p=0$ and $p=k$. Each anomalous dimension decreases at first
and later increases slowly with increasing $N_f$. Nevertheless, all
anomalous dimensions remain small. In particular, the graviton and the
fermion anomalous dimension stay below their critical values
$\eta_h<\eta_{h_\text{crit}}=2$ and
$\eta_\psi<\eta_{\psi_\text{crit}}=1$, respectively.

In summary, we find an attractive \UV{} fixed point for all numbers of
fermions. Thus, all numbers of fermions are compatible with the
asymptotic safety scenario. We also note that, in contradistinction to
the scalar case, the anomalous dimensions stay sufficiently small even
for a large number of fermions. However, the appearance of large critical
exponents is seen as an indicator for the necessity to improve the truncation.
In consequence, the impact of higher order operators will be studied in future work.

As in the scalar case we find that the $N_f$-dependence of the
couplings shown in \autoref{fig:NF_findiff_w_eta} is qualitatively
different from that in
\cite{Percacci:2002ie,Dona:2013qba,Dona:2014pla}. A detailed
comparison and evaluation of the reliability of the corresponding
approximations is deferred to \autoref{sec:background-field-approx}.

\subsection{Mixed Scalar-Fermion Systems}
\begin{figure}
 \includegraphics{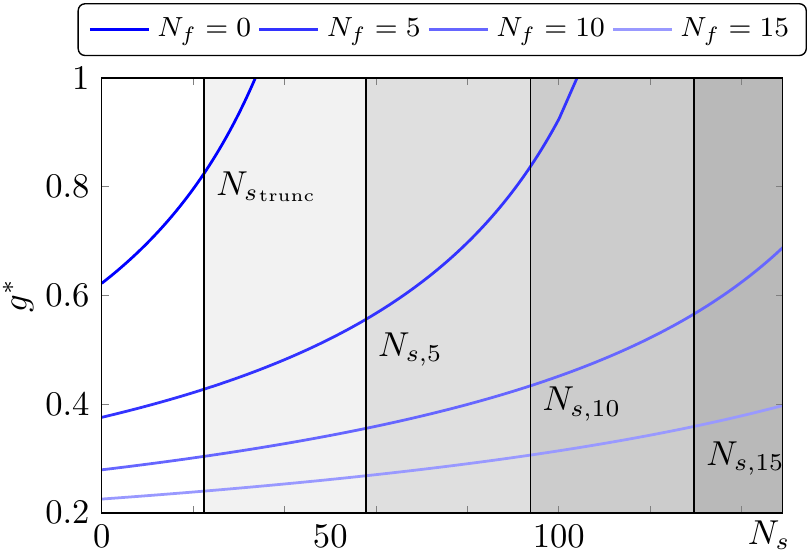}
 \caption{Fixed point value $g^*$ as a function of $N_s$ for
   $N_f=0,5,10,15$. The vertical lines denote the numbers of scalars
   for which the graviton anomalous dimension exceeds its
   critical value in the \UV{}, for the respective number of
   fermions.}\label{fig:NS_Nf}
\end{figure}
In this section, we consider the fixed point behaviour of mixed
systems of scalars and fermions. The gravity-fermion system is stable
for all $N_f$ in the present approximation. In turn, the
gravity-scalar system exceeds the bounds \eq{eq:anomdim_constr} far
before the fixed point first becomes unstable and finally
disappears. Thus, it is interesting to study the effect of a fixed
number of fermions on the $N_s$-regime of validity. As discussed in
\autoref{sec:scalars}, there exists a finite number of scalars
$N_{s_\text{trunc}}$ for which $\eta_h$ exceeds its critical value. In
\autoref{sec:fermions} we observed that the inclusion of fermions
leads to a decrease of $g^*$, which results in smaller anomalous
dimensions. Therefore, we expect that the $N_s$-regime of validity is
extended if we increase $N_f$.

In \autoref{fig:NS_Nf} the fixed point value $g^*$ is plotted as a
function of $N_s$ for different numbers of $N_f$. The vertical lines
denote the numbers of scalars for which the graviton anomalous
dimension exceeds its critical value in the \UV{}. As displayed in the
figure, the expected behaviour for the combined systems is indeed
realised. Thus, the increase of $N_f$ lowers the fixed point value
$g^*$ and extends the $N_s$-regime of validity. For $N_f=0,\,5,\,10$
and $15$ the corresponding critical values
$N_{s_\text{trunc}},N_{s,5},N_{s,10}$ and $N_{s,15}$ are given by
21.5, 57.9, 93.8 and 129.6, respectively. The maximum number of
scalars that defines the validity of the truncation increases almost
linearly with $N_f$. Thus, every additional fermion stabilises the
combined system such that $\approx7.1$ additional scalars are
admitted. The ratio between these numbers suggests that fermions have
a significantly stronger impact on the system than scalars. This is
true for the complete truncation analysed here and can also be
verified in the analytic equations by comparing the numerical values
of the respective contributions (compare \eqref{eq:ns_flows} and
\eqref{eq:nf_flows}). This imbalance between scalars and fermions was
also observed in \cite{Dona:2013qba}.  The increase of $N_f$ also
shifts the values of $N_{s_\text{sign}}$ and $N_{s_\text{max}}$ to
larger values and extends the $N_s$-regime where a fixed point is
found considerably. In summary, the inclusion of fermions stabilises
the system and extends the $N_s$-regime of validity for the given
truncation significantly.

\subsection{Independence on the approximation in the gravity sector}
We close this section with a brief discussion of the impact of the 
approximation in the pure gravity sector on our results.
Interestingly, the results agree qualitatively for 
all approximations in the pure gravity sector used in the literature. This
includes the standard ones in the background field approximation which
are discussed in the next section. We also note that the fixed
point for our truncated system is also present, if all anomalous
dimension are set to zero. It is interesting to note, however, that
for $N_s>0$, $N_f=0$ the fixed point vanishes already for $N_s \approx
45$ and therefore earlier than with anomalous
dimensions. Thus, the anomalous dimensions stabilise the \UV-behaviour
of the system.

In order to combine the present matter contributions with the pure
gravity systems in the geometrical framework \cite{Donkin:2012ud}, and
with \cite{Christiansen:2014raa}, we have to identify
$\lambda_3=\lambda_2\equiv-\mu/2$. We find that the
matter-contributions admit \UV{} fixed points. Furthermore, we observe
the same generic effect of scalars and fermions on the \UV{} fixed
point that was found for the present truncation. Hence, scalars drive
the fixed point to larger values of $g^*$. while fermions
lead to a decrease of $g^*$ and $\mu^*$,
where $\mu^*$ approaches $-1$.
In summary, our qualitative results are insensitive to 
the approximation in the pure gravity sector.

\section{Background couplings and Background-Field
  Approximation} \label{sec:background-field-approx}
It is left to study the stability of the results under a change of the
approximation scheme in the matter sector. This is even more important
as the $N_s$- and $N_f$-dependencies of the couplings shown in
\autoref{fig:NS_findiff_w_eta} and \autoref{fig:NF_findiff_w_eta} are
qualitatively different from those in
\cite{Percacci:2002ie,Dona:2013qba,Dona:2014pla}. The latter works use the
background field approximation for the computation of the flows for the couplings,
which are augmented with dynamical anomalous dimensions in
\cite{Dona:2013qba,Dona:2014pla}.  Hence, we compare the present
system of dynamical couplings with the standard flows in the background field
approximation. 

In perturbatively renormalisable quantum field theories, like the
Standard Model, the gauge invariant background couplings in the limit
$k\to{}0$ directly enter $S$-matrix computations. For $k\to0$ the
regulator, which typically depends on the background field, vanishes.
For these reasons, these couplings are observables of the
theory.  In direct analogy, we call the diffeomorphism-invariant
background couplings of quantum gravity also observables in the limit
$k\to{}0$.  Note that these quantities have a clear physical
interpretation only in the limit $k\to{}0$.  For $k>0$, on the other
hand, the background couplings depend inherently on the
background-field content via the non-vanishing regulator. In this
case, the couplings lose their clear physical meaning and their
relation to observable quantities becomes unclear
\cite{Christiansen:2015rva,Donkin:2012ud}. 

In this section we use the notation
$(g,\lambda_2,\lambda_3)$ for the dynamical couplings, where we
reintroduced $\lambda_2=-1/2\,\mu$. We also give a brief summary of
the discussion in
\cite{Pawlowski:2002eb,Litim:2002ce,Litim:2002hj,Pawlowski:2003sk,%
  Pawlowski:2005xe,Folkerts:2011jz,Donkin:2012ud,Bridle:2013sra,Dietz:2015owa}
on dynamical and background flows and the impact on the background
field approximation: Standard approaches based on diffeomorphism
invariant truncations use the background-field formalism for the
definition of the truncated effective action. The corresponding flow
equation, however, is not closed since it depends on the dynamical
propagator. This is expressed schematically as
\begin{align}\label{eq:bg_1}
  \dot \Gamma_k[\bar g, h] = F\left[\frac{\delta^2\Gamma_k[\bar
      g,h]}{\delta h^2};\bar g\right]\,,
\end{align}
where the separate dependence on $\bar g$ stems from the regulator. 
In order to close \eqref{eq:bg_1} the background-field approach
amounts to the identification of the propagators of fluctuating and
background-fields, i.e.,
\begin{align}\label{eq:bg_2}
  \frac{\delta^2\Gamma_k[\bar g,h]}{\delta
    h^2}\approx\frac{\delta^2\Gamma_k[\bar g, h]}{\delta {\bar g}^2}\,.
\end{align}
The latter identification in known to pose severe problems in \QCD,
for more details see
\cite{Folkerts:2011jz,Christiansen:2014raa}. However, at least for
pure quantum gravity the approximation \eqref{eq:bg_2} seems to work
rather well, leading to a reliable \UV-behaviour of the theory. In the
more elaborate geometrical-effective action approach
\cite{Vilkovisky:1984st,DeWitt:1988dq}, the differences between
fluctuating and background propagators are encoded in the (modified)
Nielsen-Identities \cite{Pawlowski:2003sk,Pawlowski:2005xe}. In
\cite{Donkin:2012ud} the latter identities together with a minimally
consistent extension to the Einstein-Hilbert truncation were used to
derive flow equations for the dynamical couplings $(g,\lambda)$ and
the background couplings $(\bar g,\bar \lambda)$ in the absence of
matter. In the geometrical approach the flow equations for the
background couplings read schematically
\begin{align}\label{eq:bg_flows}
  \partial_t\left(\frac{k^2}{\bar
      g}\right)=F_{R^1}(g,\lambda;N_s,N_f)\,,
  \notag\\
 \partial_t\left(\frac{\bar \lambda k^4}{\bar
     g}\right)=F_{R^0}(g,\lambda;N_s,N_f)\,,
\end{align}
for a theory with $N_s$ scalars and $N_f$ fermions. Note, that
the right hand side of the latter equation only contains dynamical
couplings. The dimensionful functions $F_{R^1}$ and $F_{R^0}$
correspond to the $R^1$ and $R^0$-terms of the required heat-kernel
expansion, respectively. With the identification of background and
dynamical couplings $(g,\lambda)=(\bar g,\bar \lambda)$, one retains
the background-field approximation from the geometrical approach.
Applying the derivatives in \eqref{eq:bg_flows} leads us to
\begin{align}\label{eq:bg_flows_2}
  \frac1{{\bar g}}\left(2-\frac{\partial_t {\bar g}}{\bar g}\right)
  =f_{R^1}(g,\lambda;N_s,N_f)\,,	\notag\\
  \frac{\bar \lambda}{\bar g}\left(4+\frac{\partial_t{\bar
        \lambda}}{\bar \lambda}-\frac{\partial_t{\bar g}}{\bar
      g}\right)=f_{R^0}(g,\lambda;N_s,N_f)\,,
\end{align}
where $f_{R^i}:=F_{R^i}\, k^{2(i-2)}$ is dimensionless. The equations
\eqref{eq:bg_flows_2} are now used to compare our flows for the
dynamical couplings $(g,\lambda_2,\lambda_3)$ with the standard
background-field flows. Since both the standard background-field
approximation and the geometrical effective action approach are based
on diffeomorphism invariant truncations, they do not distinguish
between the couplings of different-order graviton vertices. Hence, for
the present analysis we set $\lambda_3\equiv\lambda_2$ and identify
the remaining couplings $(g,\lambda_2)$ with the running dynamical
gravitational coupling and the dynamical cosmological constant in the
geometrical approach, $(g,\lambda)=(g,\lambda_2)$. We extract the
expressions for $f_{R^1}$ and $f_{R^0}$ from the flow equations in
\cite{Codello:2008vh,Dona:2013qba} reversing the identification of
background and dynamical couplings. Explicit expressions for $f_{R^i}$
are given in Appendix \ref{app:fs}.

In order to determine the fixed points of the flows
\eqref{eq:bg_flows_2}, we set $\partial_t{\bar
  g}=\partial_t{\bar\lambda}=0$ and evaluate $f_{R^i}$ at our fixed
point values for the dynamical couplings, $(g^*,\lambda_2^*)$. This
way, we arrive at simple fixed point equations for the background
couplings, to wit
\begin{align}\label{eq:bg_flows_fp}
  \bar g^*=&\frac2{f_{R^1}(g^*,\lambda^*_2;N_s,N_f)}\,	\notag\\
  \bar \lambda^*=& \frac{f_{R^0}(g^*,\lambda^*_2;N_s,N_f)}{2
    f_{R^1}(g^*,\lambda^*_2;N_s,N_f)}\,.
\end{align}
The fixed points provided by the latter equations are compared to the results from flows in the standard
background-field approximation
\cite{Codello:2008vh,Dona:2013qba}.
\begin{figure*}[t]
\centering
 \includegraphics[width=.99\textwidth]{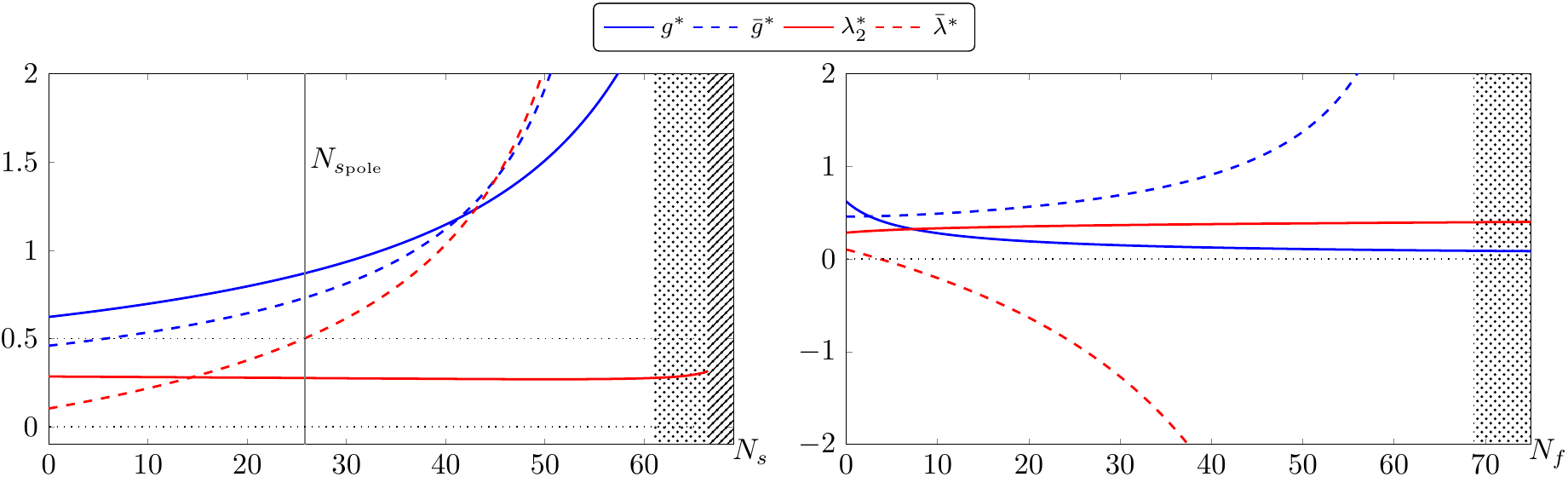}
 \caption{Fixed point values of the dynamical couplings
   $(g,\lambda_2)$ (solid lines) in comparison with their
   corresponding background counterparts $(\bar g,\bar \lambda)$
   (dashed lines) as functions of the number of scalars $N_s$ (left
   panel) and the number of fermions $N_f$ (right panel). The fixed
   point values of the background couplings diverge for $N_s=60.8$
   (left panel) and $N_f=68.6$ (right
   panel). The dotted regions denote the regimes beyond the latter
   divergences. $N_{s_\text{pole}}$ denotes the number
   of scalars at which $\bar\lambda$ would run into the propagator
   pole, if the identification $\lambda_2=\bar\lambda$, common in the
   background approximation, is applied.}
   \label{fig:BG_1}
\end{figure*}
First of all, we note that the matter-terms in the flows of the
dynamical couplings $(g,\lambda_2)$ have opposite signs
relative to the respective contributions to the flows of background
couplings. This can be seen most easily in
the analytical equations with $\vec\eta_\phi=0$ where the matter contributions to $(g,\lambda_2)$
can be written as
\begin{align}\label{eq:our_matter}
  \partial_t g \sim&-\0{43}{570 \pi} g^2 N_s
  -\0{3599}{11400\pi} g^2  N_f\,,\notag\\
  \partial_t \lambda_2 \sim&  - \01{24 \pi} g N_s+  \04{9 \pi}  g N_f\,.
\end{align}
In \cite{Codello:2008vh,Dona:2013qba} the contributions to the flows
of the background couplings ${\bar g}$ and $\bar\lambda$ read
\begin{align}\label{eq:their_matter}
  \partial_t{\bar g} \sim& +\frac1{6\pi}{\bar g}^2 N_s
  +\frac1{3\pi}\bar g^2 N_f	\notag\\
  \partial_t{\bar \lambda} \sim& +\frac1{12\pi}(3+2{\bar\lambda}){\bar
    g}N_s-\frac1{3\pi}(3-\bar\lambda){\bar g}N_f\,.
\end{align}
For $\bar\lambda<3$ every single term in \eqref{eq:our_matter} and
\eqref{eq:their_matter} carries the respective opposite sign. 

Still, the signs of the matter contributions for the background flows
are trivially the same. Accordingly, we expect the explicit $N_s, N_f$
scalings in the flows of the background couplings to dominate the
qualitative behaviour of the background fixed points. The implicit
dependence of the fixed points $(g^*,\lambda_2^*)$ on $N_s, N_f$ is
expected to be sub-leading, resulting in a similar behaviour of the
fixed points of our background quantities and those from studies in
background-field approximation.

\subsection{Background fixed points in the full system}
The left panel in \autoref{fig:BG_1} shows the fixed point for the
dynamical quantities $(g,\lambda_2)$ (solid lines) and that of their
corresponding background counterparts $(\bar g,\bar\lambda)$ (dashed
lines) calculated from \eqref{eq:bg_flows_fp} as a function of $N_s$.
The fixed point values of the background couplings have similar values
compare to the fixed points for the dynamical couplings at
$N_s=0$. However, both quantities evolve very differently under the
inclusion of scalars. In particular, $\bar g$ and $\bar \lambda$ increase
quickly with increasing $N_s$. At $N_{s_\text{pole}}=25.8$,
$\bar\lambda$ crosses the propagator pole, which is impossible in the
background-field approximation. Here, however, we do not identify
background and dynamical couplings, i.e.\ $\bar\lambda\neq\lambda_2$,
and in consequence crossing of the pole does not pose a problem. The
background couplings diverge for $N_s=60.8$, resulting
in an invalid fixed point for $N_s>60.8$ (dotted area) . The latter
divergence, however, is \textit{not} present for the dynamical
couplings. It merely results from the fact, that $f_{R^1}$ becomes
zero at this point, leading to divergent expressions for $(\bar
g,\bar\lambda)$ in \eqref{eq:bg_flows_fp}. Consequently, the fixed
point for the background couplings does in fact exist beyond
$N_s=60.8$ until the dynamical fixed point is lost (hatched
area). Since $f_{R^1}$ has, however, changed sign in this regime $\bar
g^*$ is negative and, therefore, clearly unphysical.

\begin{figure*}[t]
\centering
 \includegraphics[width=.99\textwidth]{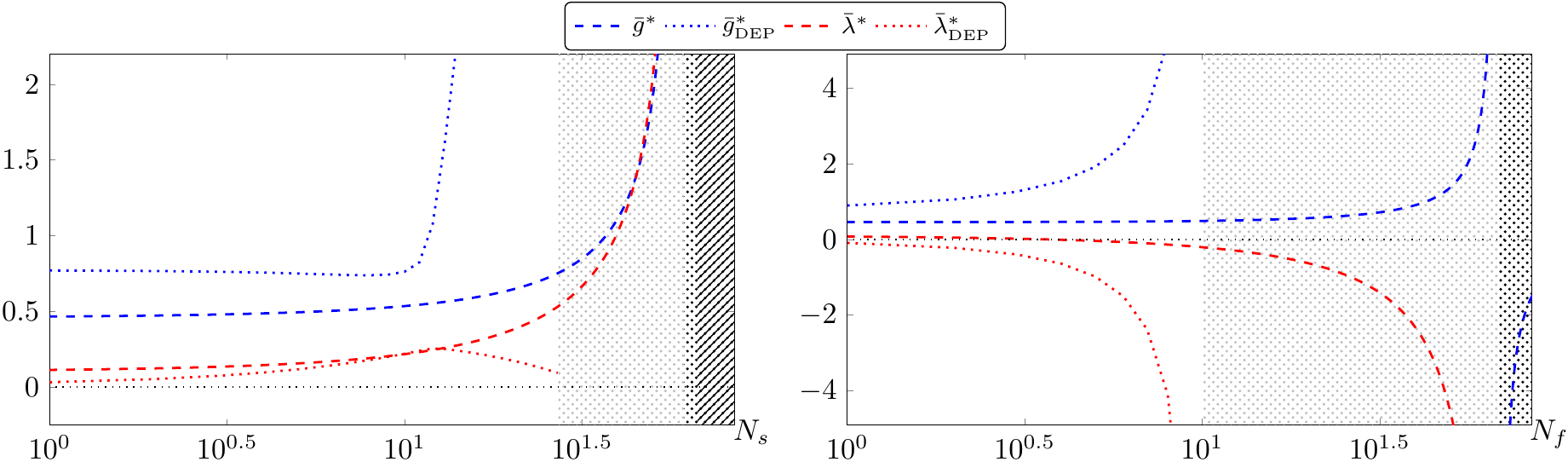}
 \caption{Logarithmic plot of the fixed point values of the background
   field couplings $(\bar g,\bar\lambda)$ as a function of the number
   of scalars $N_s$ (left) and as a function of the number of
   fermions $N_f$ (right) in comparison with results in background-field
   approximation from \cite{Dona:2013qba} (\DEP{}). Our background 
   couplings behave very similarly to the couplings in \cite{Dona:2013qba}.
   The grey and black-shaded area denote the regimes, where the fixed 
   point for the couplings in \cite{Dona:2013qba} and for our couplings 
   is lost, respectively.}
   \label{fig:BG_2}
\end{figure*}
The right panel in \autoref{fig:BG_1} compares the fixed points for
the dynamical couplings and the background couplings as a function of
$N_f$. Starting at similar values at $N_f=0$, the fixed point for the
background couplings again exhibits a very different behaviour from
that of the corresponding dynamical fixed points under the inclusion
of fermions. While $g^*$ decreases with increasing $N_f$, $\bar
g$ increases strongly. Similarly, $\bar \lambda^*$ is quickly driven
to large negative values, changing sign at $N_f=3.7$, whereas the
dynamical $\lambda_2^*$ remains almost constant. The fixed point for
the background quantities diverges for $N_f=68.6$. The dotted region
denotes the regime where the background fixed point is invalid. Again,
the divergence appears only for the background quantities. The
dynamical couplings remain well behaved for all $N_f$.

In summary, the fixed points for the background couplings behave very
differently from their dynamical counterparts under the inclusion of
matter fields. In particular, the latter exhibit divergences which are
not present for the dynamical couplings. The dynamical couplings
calculated in this work are the ones which are relevant for probing
the consistency of gravity as a quantum field theory in the
\UV{}. Thus, the above analysis suggests that divergences or the
disappearance of fixed points for the background couplings do not
reflect actual divergences of the dynamical couplings. It is therefore
indispensable, to distinguish between background and dynamical
couplings in order to study the \UV{} behaviour of quantum gravity,
once matter fields are included.

\subsection{Comparison to background fixed points in the literature}
We now compare the fixed points for the background quantities, that we
obtained from the equations \eqref{eq:bg_flows_fp}, with the ones obtained
from a background-field approximation as reported in
\cite{Dona:2013qba}. 
In our analysis, we disregard the
use of different regulators in the different approaches. Hence, we
assume that the generic behaviour of the approaches is independent of
this choice.

The left panel in \autoref{fig:BG_2} depicts fixed points for
background couplings as functions of $N_s$.  The dotted curves
represent fixed points of flows determined in background-field
approximation in \cite{Dona:2013qba} (\DEP{}) and the dashed curves
denote our background couplings, which are calculated from the
dynamical couplings (identical to the respective curves in
\autoref{fig:BG_1}).  The fixed point value for the gravitational
coupling $\bar g^*_\text{DEP}$ increases with increasing $N_s$ and
eventually diverges at $N_s\approx27$. For $N_s>27$ no \UV{} fixed
point exists, which is indicated by the grey dotted area in the plot.
Note, that due to the identification of background and dynamical
couplings the graviton-propagator pole is located at $\bar
\lambda=0.5$. This limit cannot be intersected by
$\bar\lambda^*_\text{DEP}$. In consequence, $\bar\lambda^*_\text{DEP}$
first increases but exhibits a characteristic kink at $N_s\approx16$
and then decreases again until the fixed point ceases to exist at
$N_s\approx27$.

For small numbers of $N_s$, the fixed points from the background-field
approach $(\bar g^*_\text{DEP},\bar \lambda^*_\text{DEP})$ show a
behaviour, which is similar to that of our background couplings $(\bar
g^*,\bar \lambda^*)$. For larger values of $N_s$ the value of
$\bar\lambda^*_\text{DEP}$ is driven closer to the propagator pole and
the flow equations receive growing contributions from the graviton
loops, which is not the case for our background couplings. Here, the
implicit dependence of the fixed point on $N_s$ is large and we
observe large deviations between our background couplings and those in
\cite{Dona:2013qba} in the regime $N_s\gtrsim10$.

The right panel in \autoref{fig:BG_2} depicts the fixed points for
background couplings as functions of $N_f$. The notation for the
curves in the right panel is the same as in the left one described
above. The fixed point value $\bar g^*_\text{DEP}$ strongly increases
with increasing $N_f$ and runs into a divergence for
$N_f\approx10$. For $N_f>10$ the fixed point does not exist anymore,
which is indicated by the grey dotted area in the plot.  The value for
$\bar\lambda^*_\text{DEP}$ starts at small negative values and
decreases quickly with increasing $N_f$ until the fixed point ceases
to exist at $N_f\approx10$.

For small numbers of $N_f$, our background couplings $(\bar
g^*,\bar\lambda^*)$ show again a similar behaviour to the fixed points
in the background-field approximation. For larger $N_f$, we observe
large deviations, though the generic behaviour of the fixed points is
the same.  An important common feature is the existence of a
singularity for the fixed point for a finite number of fermions.  As
discussed in the previous section, this divergence has no influence on
the asymptotic safety of the theory since it is clearly independent
from the physical dynamical couplings. Again, the divergence of the
background fixed point is due to the fact that $f_{R^1}$ in
\eqref{eq:bg_flows_fp} passes zero. Beyond this divergence the
background fixed point still exists but has changed sign. This can be
observed for $\bar g^*$ in the lower right corner of the right panel
of \autoref{fig:BG_2}.

In summary, for sufficiently small $N_f,N_s \lesssim 10$ the couplings
in the background-field approximation (\DEP) behave similarly to the
background-field couplings of the full dynamical system computed
here. Note, that both computations show divergences in the background
coupling for a finite number of scalars and fermions. These
divergences are not reflected in the dynamical couplings and the
current analysis strongly suggests their absence at $k=0$. We conclude
that the background-field approximation provides an adequate
qualitative picture of the behaviour of the physical background
couplings for $N_f,N_s \lesssim 10$. The relevant quantities for
studies of the \UV{} behaviour of quantum gravity are, however, the
dynamical couplings.  In turn, for $N_f,N_s \gtrsim 10$ the background
field approximation fails, and it is necessary to compute dynamical
flows and couplings.

\section{Summary}
We have presented the first genuine calculation of dynamical
gravitational couplings based on a vertex flow in gravity-matter
systems with an arbitrary number of scalars and fermions.  We
have calculated the matter contributions to the dynamical graviton
two- and three-point functions and included momentum-dependent gravity
and matter anomalous dimensions. The \UV{} behaviour of the resulting
theory has been analysed under the influence of $N_s$ scalars and $N_f$ fermions. 

In the scalar sector the increase of $N_s$ leads to an increasing
Newton's coupling at the \UV{} fixed point and thus to a strengthening
of graviton fluctuations at high energies. For large numbers of
scalars $N_s>21.5$ the present generic class of regulators violates
the bounds \eq{eq:anomdim_constr} due to a large graviton anomalous
dimension, i.e.\ $\eta_h>2$ in this regime. Deep in this regime the
\UV{} fixed point first becomes repulsive and finally is lost, which
requires further investigation.

In the fermion sector the \UV{} fixed point exists and is stable for
all $N_f$. Also, all fixed point values remain small, and the
anomalous dimensions stay below the bounds \eq{eq:anomdim_constr},
i.e.\ $\eta_h,\eta_c,\eta_\varphi <2$ and $\eta_\psi<1$, for all
$N_f$.  Similar to the scalar case the increase of $N_f$ enhances
graviton fluctuations. Here however, the enhancement is due to the
shift of the graviton-mass parameter towards the propagator pole.

In summary, we always find an attractive \UV{} fixed point in the
presence of a general number of scalars and fermions within the
validity bounds for the generic class of regulators used here. Finally
we have discussed and embedded previous results in the literature
within our extended setting. In particular we have also compared the
present results within the full dynamical system to results that
partially rely on the background field approximation. Interestingly,
we find the signs of the matter contributions to the flows of our
dynamical couplings to be opposite to those of flows in
background-field approximation. This is in sharp contrast to the pure
gravity flows whose signs agree in all approximations. We have also
computed the fixed points of the background couplings in the present
approach. We have shown that the latter agree qualitatively with the
fixed point couplings in in the background-field approximation for
$N_f,N_s \lesssim 10$.  In turn, for $N_f,N_s \gtrsim 10$ the
background field approximation fails, and it is necessary to compute
dynamical flows and couplings.

Currently, we extend the present work to vector fields as well as to
improved approximations. This includes approximations that are not
sensitive to the validity bounds for the generic class of regulators
used here related to the size of the anomalous dimensions as well as
including higher orders in the curvature scalar $R$.\\[-1.5ex]

\noindent {\bf Acknowledgements} 
We thank N.~Christiansen, A.~Eichhorn, K.~Falls, H.~Gies, T.~Henz, R.~Percacci, 
A.~Rodigast and C.~Wetterich for discussions. MR acknowledges funding
from IMPRS-PTFS. This work is supported by EMMI and by ERC-AdG-290623.

\appendix
\section{Flow of the Three-Point Function}\label{app:three-point-function}
In \autoref{sec:frg} and in \autoref{sec:gravity-flows} we discussed
the projection and the flow of the graviton three-point function. Here
we provide the explicit form of the projection operators, and the
corresponding projected flow equations.  In the three-step procedure
outlined in \autoref{sec:frg} we presented a way to construct two
projection operators for the three-point function. From the momentum-independent
part of $\mathcal{T}^{(3)}$ we obtained a projection operator for
$\partial_t \Lambda_3$, which we call $\Pi_{\Lambda_3}$.  In turn, the
corresponding projection operator for $\partial_t G$, $\Pi_{G}$, was constructed
from the momentum dependent part of $\mathcal{T}^{(3)}$. In
a multi-index notation, $\Pi_{G}$ (in the symmetric momentum
configuration) and $\Pi_{\Lambda_3}$ are given by
\begin{align}
  \Pi_G^{ABC} &= \Pi_\TT^{AA'}(p_1^2) \Pi_\TT^{BB'}(p_2^2)
  \Pi_\TT^{CC'}(p_3^2)
  \0{\mathcal{T}^{(3)}_{A'B'C'}(\mathbf{p};0)}{p_1^2}\,,  \label{eq:pi_g} \\
  \Pi_{\Lambda_3}^{ABC} &= \Pi_\TT^{AA'}(p_1^2) \Pi_\TT^{BB'}(p_2^2)
  \Pi_\TT^{CC'}(p_3^2)\mathcal{T}^{(3)}_{A'B'C'}(0;1)
  \,,\label{eq:pi_lam}
\end{align}
where $A$, $B$ and $C$ are multi-indices, e.g.\
$A=\mu\nu$. Contracting $\Gamma^{(hhh)}_k$ with these objects leads to 
scalar expressions, which we call $\Gamma^{(hhh)}_{\TT,G}$ and
$\Gamma^{(hhh)}_{\TT,\Lambda}$, respectively. The latter are given
schematically in equation \eqref{eq:Gamma_3h} and read explicitly
\begin{align} \label{eq:Gamma_3h-full} \Gamma^{(hhh)}_{\TT,G}(p^2)&=
  G^{1/2}\0{Z_h^{3/2}(p^2)}{(32\pi)^2}
  \left(\0{171}{32} p^2 - \094 \Lambda_3\right)\,, \\
  \Gamma^{(hhh)}_{\TT,\Lambda}(0)&= G^{1/2}\0{Z_h^{3/2}(0)}{(32\pi)^2}
  \0{80}{3}\Lambda_3\,, \label{eq:Gamma_3h-full-Lambda}
\end{align}
where the subscript $G$ and $\Lambda$ refer to the different
projections schemes as described in the equations \eqref{eq:pi_g} and \eqref{eq:pi_lam}. From
these equations we take a scale derivative and divide by the
appropriate wavefunction renormalisation, i.e.\ $Z_h^{3/2}(p^2)$ for
equation \eq{eq:Gamma_3h-full} and $Z_h^{3/2}(0)$ for equation
\eq{eq:Gamma_3h-full-Lambda}. Afterwards \eq{eq:Gamma_3h-full} is evaluated at
$p=k$ as well as $p=0$. The respective results are subtracted from
each other. With the usual dimensionless quantities introduced in
\eqref{eq:dimless} this leads to the flow equations
\begin{align}\label{eq:flow-3h-full}
  \dot g =& 2g + 3 \eta_h(k^2) g -\0{24}{19}\left(\eta_h(k^2)
    -\eta_h(0)\right)\lambda_3\ g \notag\\ &+ \0{64}{171}
  \frac{(32\pi)^2\sqrt{g}}{k}\left(\text{Flow}^{(hhh)}_{\TT,G}(k^2)-
    \text{Flow}^{(hhh)}_{\TT,G}(0)\right)\,, \\
  \dot \lambda_3 =& -2\lambda_3 + \032\eta_h(0)\lambda_3
  +\012\left( 2g-\dot g \right)\frac{\lambda_3}{g} \notag\\
  &+
  \0{3}{80}\0{(32\pi)^2}{\sqrt{g}k}\text{Flow}_{\TT,\Lambda}^{(hhh)}(0)\,.
\end{align}
Note, that prefactors such as $\0{24}{19}$ or $\0{64}{171}$ depend on the
kinematic configuration. The present flow equations are evaluated for
the symmetric momentum configuration, see
\eqref{eq:mom-config}. The prefactors in front of $\text{Flow}$ also
depend on the norm of the projection operators. The present numbers
are obtained with unnormalised transverse-traceless projection operators, i.e.\
$\Pi_\TT\circ\Pi_\TT=5$. These equations do not have an analytic
form. To obtain analytic equations, a derivative projection is
necessary, but this is less accurate in capturing the momentum
dependence of the flow, see Appendix \ref{app:analytic-eq}.

\section{Anomalous Dimensions}\label{app:anom_dim}
The anomalous dimensions obey a system of coupled Fredholm integral
equations. The latter is given by
\begin{align}
  \eta_h(p^2) &= 32\pi\0{\text{Flow}_{\TT}^{(hh)}(-M^2)-
    \text{Flow}_{\TT}^{(hh)}(p^2)}{p^2+M^2}[\vec \eta_\phi]\,, \notag\\
  \eta_c (p^2) &= -\0{\text{Flow}^{(\bar c c)}(p^2)}{p^2}[
  \eta_h,\eta_c]\,,\notag\\
  \eta_\psi (p^2) &= i\0{\text{tr}(\slashed p
    \text{Flow}^{(\bar\psi\psi)})(p^2)}{ d \,p^2}[\eta_h,\eta_\psi]\,, \notag\\
  \eta_\varphi (p^2) &= -\0{\text{Flow}^{(\varphi\varphi)}(p^2)}{p^2}[\eta_h,\eta_\varphi]\,.
 \label{eq:eq-anom-dim}
\end{align}
The squared brackets denote functional dependencies on the respective
anomalous dimensions. The content of the brackets also indicates which
fields run in the loop of corresponding two-point function. We
approximate the equations \eqref{eq:eq-anom-dim} by evaluating the
anomalous dimension at the momentum scale $p=k$, see
\autoref{sec:anom-dim}.

\section{Analytic Flow Equations}\label{app:analytic-eq}
Throughout this work we have used the full numerical flow equations to
compute the \UV{} fixed points. Nevertheless, we derived analytic flow
equations, which are, however, less accurate in capturing the momentum
dependence of the flow \cite{Christiansen:2015rva}. To obtain analytic
flow equations we need to employ
\begin{itemize}
\setlength{\itemsep}{10pt}
 \item a Litim-type regulator,
 
 \item the momentum approximation of the anomalous dimension from
   \autoref{sec:anom-dim} in each loop integral,
 
 \item a derivative projection for $\partial_t g$ instead of the usual
   bilocal projection (for bilocal projection see Appendix
   \ref{app:three-point-function}).
\end{itemize}
The latter implies the following: As usual, we take a scale derivative
of equation \eqref{eq:Gamma_3h-full} and divide by $Z_h^{3/2}(p^2)$
and $p^2$. Then, we take another derivative, this time with respect to
$p^2$ and evaluate the result at $p=0$. Now, the loop integration can
be performed analytically. The resulting analytic equations are
\begin{widetext}
\begin{align}
  \dot g=&\left(2+3\eta_h(0)-\0{24}{19}\eta_h'(0)\lambda_3\right)g\notag\\
  &+\0{g^2}{\pi}\left(-\0{47 (6-\eta_h(k^2))}{114 (\mu+1)^2} +
    \0{\left(472(6-
        \eta_h(k^2))-360(4-\eta_h(k^2))\lambda_3\right)\lambda_4-240(6-\eta_h(k^2))
      \lambda_3+45(8-\eta_h(k^2))}{342 (\mu+1)^3}\right.\notag\\
  &\left.\hspace*{1.1cm}+ \0{16(1-3\lambda_3)\lambda_4}{19(\mu+1)^4} +
    \0{25920(4-\eta_h(k^2))\lambda_3^3+3380(6-\eta_h(k^2))\lambda_3^2-
      1860(8-\eta_h(k^2)) \lambda_3+147 (10-\eta_h(k^2))}{1710(\mu+1)^4}\right.\notag\\
  &\left.\hspace*{1.1cm}+ 2\0{2336 \lambda_3^3-3640 \lambda_3^2 +1780
      \lambda_3-299}{285 (\mu+1)^5} - \0{53 (10-\eta_c(k^2))}{190}+
    \0{48}{19}\right)\notag\\
  &+N_f \0{g^2}{\pi} \left(-\0{521(6-\eta_\psi(k^2))}{17100}-\0{3(5-
      \eta_\psi(k^2))}{152}-\0{13}{380}\right) \notag\\
  &+N_s \0{g^2}{\pi} \left(\0{10-\eta_\varphi(k^2)}{1140}-\08{95}\right)\,, \notag\\[1.5ex]
  \dot \lambda_3=&\left(-1+\032\eta_h(0)-\0{\dot g}{2g}\right)\lambda_3\notag\\
  &+\0g{\pi} \left(\0{8-\eta_h(k^2)-4 (6-\eta_h(k^2))
      \lambda_5}{8(\mu+1)^2} +
    \0{(-16(6-\eta_h(k^2))\lambda_3+3(8-\eta_h(k^2))) \lambda_4}{6(\mu
      +1)^3}\right.\notag\\
  &\left.\hspace*{0.97cm}+ \0{80(6-\eta_h(k^2))\lambda_3^3-120(8-
      \eta_h(k^2))\lambda_3^2+72(10-\eta_h(k^2))\lambda_3-11(12-
      \eta_h(k^2))}{240(\mu+1)^4} + \0{12-\eta_c(k^2)}{10}\right)\notag\\
  &+ N_f \0g{\pi}\left(\0{8-\eta_\psi(k^2)}{224}-\0{7-
      \eta_\psi(k^2)}{56}+\0{17 (6-\eta_\psi(k^2))}{240}\right)\notag\\
  &+ N_s \0g{\pi} \left(\0{12-\eta_\varphi(k^2)}{480}-
    \0{10-\eta_\varphi(k^2)}{80}+\0{8-\eta_\varphi(k^2)}{96}\right)\,, \notag\\[1.5ex]
\dot \mu=&\Big(-2+\eta_h(0)\Big)\mu\notag\\
&+\0g{\pi} \left(\0{8(6-\eta_h(k^2))\lambda_4-3(8- \eta_h(k^2))}{12
    (\mu+1)^2}+\0{320(6-\eta_h(k^2))\lambda_3^2
    -120(8-\eta_h(k^2))\lambda_3+21(10-\eta_h(k^2))}{
    180(\mu+1)^3} \right. \notag\\
&\left.\hspace*{0.97cm} - \0{10-\eta_c(k^2)}5\right)\notag\\
&+ N_f \0g{\pi}\left(\0{7-\eta_\psi(k^2)}{63}-\0{6-\eta_\psi(k^2)}{6}\right)\notag\\
&+ N_s \0g{\pi} \left(\0{10-\eta_\varphi(k^2)}{120}\right)\,.
\label{eq:analyticflows}
\end{align}
\end{widetext}
\section{Background Quantities}\label{app:fs}
The functions $f_{R^i}$, which are discussed in
\autoref{sec:background-field-approx}, are extracted from
\cite{Dona:2013qba}. In our case they read
\begin{widetext}
\begin{align}
  f_{R^0}(g,\lambda,N_s,N_f)=&\frac{1}{48 \pi} \left(\frac{20 (6
      -\eta_h(k^2))}{1-2 \lambda}-16 (6-\eta_c(k^2))+
    2 (6-\eta_\varphi(k^2)) N_s-8 (6-\eta_\psi(k^2)) N_f\right)\,,\notag\\
  f_{R^1}(g,\lambda,N_s,N_f)=&\frac{1}{48 \pi} \left(\frac{52
      (4-\eta_h(k^2))}{1-2 \lambda}+40 (4-\eta_c(k^2))-2
    (4-\eta_\varphi(k^2)) N_s-4 (4-\eta_\psi(k^2))
    N_f\right)\,. \label{eq:fRi}
\end{align}
\end{widetext}
In order to obtain the functions in equation \eq{eq:fRi}, we reversed
the identification of background and dynamical quantities and replaced
$\eta_\phi\to\eta_\phi(k^2)$ in order to evaluate the anomalous
dimension at the values, where the integrals are peaked. Note, that
the functions $f_{R^i}$ depend on the dynamical gravitational coupling
$g$ only via the anomalous dimensions.
\bibliography{bib_gravity-fermions}

\begin{thebibliography}{69}%
\makeatletter
\providecommand \@ifxundefined [1]{%
 \@ifx{#1\undefined}
}%
\providecommand \@ifnum [1]{%
 \ifnum #1\expandafter \@firstoftwo
 \else \expandafter \@secondoftwo
 \fi
}%
\providecommand \@ifx [1]{%
 \ifx #1\expandafter \@firstoftwo
 \else \expandafter \@secondoftwo
 \fi
}%
\providecommand \natexlab [1]{#1}%
\providecommand \enquote  [1]{``#1''}%
\providecommand \bibnamefont  [1]{#1}%
\providecommand \bibfnamefont [1]{#1}%
\providecommand \citenamefont [1]{#1}%
\providecommand \href@noop [0]{\@secondoftwo}%
\providecommand \href [0]{\begingroup \@sanitize@url \@href}%
\providecommand \@href[1]{\@@startlink{#1}\@@href}%
\providecommand \@@href[1]{\endgroup#1\@@endlink}%
\providecommand \@sanitize@url [0]{\catcode `\\12\catcode `\$12\catcode
  `\&12\catcode `\#12\catcode `\^12\catcode `\_12\catcode `\%12\relax}%
\providecommand \@@startlink[1]{}%
\providecommand \@@endlink[0]{}%
\providecommand \url  [0]{\begingroup\@sanitize@url \@url }%
\providecommand \@url [1]{\endgroup\@href {#1}{\urlprefix }}%
\providecommand \urlprefix  [0]{URL }%
\providecommand \Eprint [0]{\href }%
\providecommand \doibase [0]{http://dx.doi.org/}%
\providecommand \selectlanguage [0]{\@gobble}%
\providecommand \bibinfo  [0]{\@secondoftwo}%
\providecommand \bibfield  [0]{\@secondoftwo}%
\providecommand \translation [1]{[#1]}%
\providecommand \BibitemOpen [0]{}%
\providecommand \bibitemStop [0]{}%
\providecommand \bibitemNoStop [0]{.\EOS\space}%
\providecommand \EOS [0]{\spacefactor3000\relax}%
\providecommand \BibitemShut  [1]{\csname bibitem#1\endcsname}%
\let\auto@bib@innerbib\@empty
\bibitem [{\citenamefont {Christiansen}\ \emph {et~al.}(2015)\citenamefont
  {Christiansen}, \citenamefont {Knorr}, \citenamefont {Meibohm}, \citenamefont
  {Pawlowski},\ and\ \citenamefont {Reichert}}]{Christiansen:2015rva}%
  \BibitemOpen
  \bibfield  {author} {\bibinfo {author} {\bibfnamefont {N.}~\bibnamefont
  {Christiansen}}, \bibinfo {author} {\bibfnamefont {B.}~\bibnamefont {Knorr}},
  \bibinfo {author} {\bibfnamefont {J.}~\bibnamefont {Meibohm}}, \bibinfo
  {author} {\bibfnamefont {J.~M.}\ \bibnamefont {Pawlowski}}, \ and\ \bibinfo
  {author} {\bibfnamefont {M.}~\bibnamefont {Reichert}},\ }\href {\doibase
  10.1103/PhysRevD.92.121501} {\bibfield  {journal} {\bibinfo  {journal} {Phys.
  Rev.}\ }\textbf {\bibinfo {volume} {D92}},\ \bibinfo {pages} {121501}
  (\bibinfo {year} {2015})},\ \Eprint {http://arxiv.org/abs/1506.07016}
  {arXiv:1506.07016 [hep-th]} \BibitemShut {NoStop}%
\bibitem [{\citenamefont {Weinberg}(1979)}]{Weinberg:1980gg}%
  \BibitemOpen
  \bibfield  {author} {\bibinfo {author} {\bibfnamefont {S.}~\bibnamefont
  {Weinberg}},\ }\href@noop {} {\bibfield  {journal} {\bibinfo  {journal}
  {General Relativity: An Einstein centenary survey, Eds. Hawking, S.W.,
  Israel, W; Cambridge University Press}\ ,\ \bibinfo {pages} {790}} (\bibinfo
  {year} {1979})}\BibitemShut {NoStop}%
\bibitem [{\citenamefont {Reuter}(1998)}]{Reuter:1996cp}%
  \BibitemOpen
  \bibfield  {author} {\bibinfo {author} {\bibfnamefont {M.}~\bibnamefont
  {Reuter}},\ }\href {\doibase 10.1103/PhysRevD.57.971} {\bibfield  {journal}
  {\bibinfo  {journal} {Phys. Rev.}\ }\textbf {\bibinfo {volume} {D57}},\
  \bibinfo {pages} {971} (\bibinfo {year} {1998})},\ \Eprint
  {http://arxiv.org/abs/hep-th/9605030} {arXiv:hep-th/9605030} \BibitemShut
  {NoStop}%
\bibitem [{\citenamefont {Wetterich}(1993)}]{Wetterich:1992yh}%
  \BibitemOpen
  \bibfield  {author} {\bibinfo {author} {\bibfnamefont {C.}~\bibnamefont
  {Wetterich}},\ }\href {\doibase 10.1016/0370-2693(93)90726-X} {\bibfield
  {journal} {\bibinfo  {journal} {Phys.Lett.}\ }\textbf {\bibinfo {volume}
  {B301}},\ \bibinfo {pages} {90} (\bibinfo {year} {1993})}\BibitemShut
  {NoStop}%
\bibitem [{\citenamefont {Souma}(1999)}]{Souma:1999at}%
  \BibitemOpen
  \bibfield  {author} {\bibinfo {author} {\bibfnamefont {W.}~\bibnamefont
  {Souma}},\ }\href {\doibase 10.1143/PTP.102.181} {\bibfield  {journal}
  {\bibinfo  {journal} {Prog.Theor.Phys.}\ }\textbf {\bibinfo {volume} {102}},\
  \bibinfo {pages} {181} (\bibinfo {year} {1999})},\ \Eprint
  {http://arxiv.org/abs/hep-th/9907027} {arXiv:hep-th/9907027 [hep-th]}
  \BibitemShut {NoStop}%
\bibitem [{\citenamefont {Reuter}\ and\ \citenamefont
  {Saueressig}(2002)}]{Reuter:2001ag}%
  \BibitemOpen
  \bibfield  {author} {\bibinfo {author} {\bibfnamefont {M.}~\bibnamefont
  {Reuter}}\ and\ \bibinfo {author} {\bibfnamefont {F.}~\bibnamefont
  {Saueressig}},\ }\href {\doibase 10.1103/PhysRevD.65.065016} {\bibfield
  {journal} {\bibinfo  {journal} {Phys. Rev.}\ }\textbf {\bibinfo {volume}
  {D65}},\ \bibinfo {pages} {065016} (\bibinfo {year} {2002})},\ \Eprint
  {http://arxiv.org/abs/hep-th/0110054} {arXiv:hep-th/0110054 [hep-th]}
  \BibitemShut {NoStop}%
\bibitem [{\citenamefont {Christiansen}\ \emph {et~al.}(2014)\citenamefont
  {Christiansen}, \citenamefont {Litim}, \citenamefont {Pawlowski},\ and\
  \citenamefont {Rodigast}}]{Christiansen:2012rx}%
  \BibitemOpen
  \bibfield  {author} {\bibinfo {author} {\bibfnamefont {N.}~\bibnamefont
  {Christiansen}}, \bibinfo {author} {\bibfnamefont {D.~F.}\ \bibnamefont
  {Litim}}, \bibinfo {author} {\bibfnamefont {J.~M.}\ \bibnamefont
  {Pawlowski}}, \ and\ \bibinfo {author} {\bibfnamefont {A.}~\bibnamefont
  {Rodigast}},\ }\href {\doibase 10.1016/j.physletb.2013.11.025} {\bibfield
  {journal} {\bibinfo  {journal} {Phys.Lett.}\ }\textbf {\bibinfo {volume}
  {B728}},\ \bibinfo {pages} {114} (\bibinfo {year} {2014})},\ \Eprint
  {http://arxiv.org/abs/1209.4038} {arXiv:1209.4038 [hep-th]} \BibitemShut
  {NoStop}%
\bibitem [{\citenamefont {Christiansen}\ \emph {et~al.}(2016)\citenamefont
  {Christiansen}, \citenamefont {Knorr}, \citenamefont {Pawlowski},\ and\
  \citenamefont {Rodigast}}]{Christiansen:2014raa}%
  \BibitemOpen
  \bibfield  {author} {\bibinfo {author} {\bibfnamefont {N.}~\bibnamefont
  {Christiansen}}, \bibinfo {author} {\bibfnamefont {B.}~\bibnamefont {Knorr}},
  \bibinfo {author} {\bibfnamefont {J.~M.}\ \bibnamefont {Pawlowski}}, \ and\
  \bibinfo {author} {\bibfnamefont {A.}~\bibnamefont {Rodigast}},\ }\href
  {\doibase 10.1103/PhysRevD.93.044036} {\bibfield  {journal} {\bibinfo
  {journal} {Phys. Rev.}\ }\textbf {\bibinfo {volume} {D93}},\ \bibinfo {pages}
  {044036} (\bibinfo {year} {2016})},\ \Eprint {http://arxiv.org/abs/1403.1232}
  {arXiv:1403.1232 [hep-th]} \BibitemShut {NoStop}%
\bibitem [{\citenamefont {Donkin}\ and\ \citenamefont
  {Pawlowski}(2012)}]{Donkin:2012ud}%
  \BibitemOpen
  \bibfield  {author} {\bibinfo {author} {\bibfnamefont {I.}~\bibnamefont
  {Donkin}}\ and\ \bibinfo {author} {\bibfnamefont {J.~M.}\ \bibnamefont
  {Pawlowski}},\ }\href@noop {} {\  (\bibinfo {year} {2012})},\ \Eprint
  {http://arxiv.org/abs/1203.4207} {arXiv:1203.4207 [hep-th]} \BibitemShut
  {NoStop}%
\bibitem [{\citenamefont {Falls}\ \emph {et~al.}(2014)\citenamefont {Falls},
  \citenamefont {Litim}, \citenamefont {Nikolakopoulos},\ and\ \citenamefont
  {Rahmede}}]{Falls:2014tra}%
  \BibitemOpen
  \bibfield  {author} {\bibinfo {author} {\bibfnamefont {K.}~\bibnamefont
  {Falls}}, \bibinfo {author} {\bibfnamefont {D.~F.}\ \bibnamefont {Litim}},
  \bibinfo {author} {\bibfnamefont {K.}~\bibnamefont {Nikolakopoulos}}, \ and\
  \bibinfo {author} {\bibfnamefont {C.}~\bibnamefont {Rahmede}},\ }\href@noop
  {} {\  (\bibinfo {year} {2014})},\ \Eprint {http://arxiv.org/abs/1410.4815}
  {arXiv:1410.4815 [hep-th]} \BibitemShut {NoStop}%
\bibitem [{\citenamefont {Lauscher}\ and\ \citenamefont
  {Reuter}(2002)}]{Lauscher:2002sq}%
  \BibitemOpen
  \bibfield  {author} {\bibinfo {author} {\bibfnamefont {O.}~\bibnamefont
  {Lauscher}}\ and\ \bibinfo {author} {\bibfnamefont {M.}~\bibnamefont
  {Reuter}},\ }\href {\doibase 10.1103/PhysRevD.66.025026} {\bibfield
  {journal} {\bibinfo  {journal} {Phys. Rev.}\ }\textbf {\bibinfo {volume}
  {D66}},\ \bibinfo {pages} {025026} (\bibinfo {year} {2002})},\ \Eprint
  {http://arxiv.org/abs/hep-th/0205062} {arXiv:hep-th/0205062} \BibitemShut
  {NoStop}%
\bibitem [{\citenamefont {Codello}\ and\ \citenamefont
  {Percacci}(2006)}]{Codello:2006in}%
  \BibitemOpen
  \bibfield  {author} {\bibinfo {author} {\bibfnamefont {A.}~\bibnamefont
  {Codello}}\ and\ \bibinfo {author} {\bibfnamefont {R.}~\bibnamefont
  {Percacci}},\ }\href {\doibase 10.1103/PhysRevLett.97.221301} {\bibfield
  {journal} {\bibinfo  {journal} {Phys. Rev. Lett.}\ }\textbf {\bibinfo
  {volume} {97}},\ \bibinfo {pages} {221301} (\bibinfo {year} {2006})},\
  \Eprint {http://arxiv.org/abs/hep-th/0607128} {arXiv:hep-th/0607128}
  \BibitemShut {NoStop}%
\bibitem [{\citenamefont {Codello}\ \emph {et~al.}(2008)\citenamefont
  {Codello}, \citenamefont {Percacci},\ and\ \citenamefont
  {Rahmede}}]{Codello:2007bd}%
  \BibitemOpen
  \bibfield  {author} {\bibinfo {author} {\bibfnamefont {A.}~\bibnamefont
  {Codello}}, \bibinfo {author} {\bibfnamefont {R.}~\bibnamefont {Percacci}}, \
  and\ \bibinfo {author} {\bibfnamefont {C.}~\bibnamefont {Rahmede}},\ }\href
  {\doibase 10.1142/S0217751X08038135} {\bibfield  {journal} {\bibinfo
  {journal} {Int. J. Mod. Phys.}\ }\textbf {\bibinfo {volume} {A23}},\ \bibinfo
  {pages} {143} (\bibinfo {year} {2008})},\ \Eprint
  {http://arxiv.org/abs/0705.1769} {arXiv:0705.1769 [hep-th]} \BibitemShut
  {NoStop}%
\bibitem [{\citenamefont {Codello}\ \emph {et~al.}(2009)\citenamefont
  {Codello}, \citenamefont {Percacci},\ and\ \citenamefont
  {Rahmede}}]{Codello:2008vh}%
  \BibitemOpen
  \bibfield  {author} {\bibinfo {author} {\bibfnamefont {A.}~\bibnamefont
  {Codello}}, \bibinfo {author} {\bibfnamefont {R.}~\bibnamefont {Percacci}}, \
  and\ \bibinfo {author} {\bibfnamefont {C.}~\bibnamefont {Rahmede}},\ }\href
  {\doibase 10.1016/j.aop.2008.08.008} {\bibfield  {journal} {\bibinfo
  {journal} {Annals Phys.}\ }\textbf {\bibinfo {volume} {324}},\ \bibinfo
  {pages} {414} (\bibinfo {year} {2009})},\ \Eprint
  {http://arxiv.org/abs/0805.2909} {arXiv:0805.2909 [hep-th]} \BibitemShut
  {NoStop}%
\bibitem [{\citenamefont {Machado}\ and\ \citenamefont
  {Saueressig}(2008)}]{Machado:2007ea}%
  \BibitemOpen
  \bibfield  {author} {\bibinfo {author} {\bibfnamefont {P.~F.}\ \bibnamefont
  {Machado}}\ and\ \bibinfo {author} {\bibfnamefont {F.}~\bibnamefont
  {Saueressig}},\ }\href {\doibase 10.1103/PhysRevD.77.124045} {\bibfield
  {journal} {\bibinfo  {journal} {Phys. Rev.}\ }\textbf {\bibinfo {volume}
  {D77}},\ \bibinfo {pages} {124045} (\bibinfo {year} {2008})},\ \Eprint
  {http://arxiv.org/abs/0712.0445} {arXiv:0712.0445 [hep-th]} \BibitemShut
  {NoStop}%
\bibitem [{\citenamefont {Benedetti}\ \emph {et~al.}(2009)\citenamefont
  {Benedetti}, \citenamefont {Machado},\ and\ \citenamefont
  {Saueressig}}]{Benedetti:2009rx}%
  \BibitemOpen
  \bibfield  {author} {\bibinfo {author} {\bibfnamefont {D.}~\bibnamefont
  {Benedetti}}, \bibinfo {author} {\bibfnamefont {P.~F.}\ \bibnamefont
  {Machado}}, \ and\ \bibinfo {author} {\bibfnamefont {F.}~\bibnamefont
  {Saueressig}},\ }\href {\doibase 10.1142/S0217732309031521} {\bibfield
  {journal} {\bibinfo  {journal} {Mod. Phys. Lett.}\ }\textbf {\bibinfo
  {volume} {A24}},\ \bibinfo {pages} {2233} (\bibinfo {year} {2009})},\ \Eprint
  {http://arxiv.org/abs/0901.2984} {arXiv:0901.2984 [hep-th]} \BibitemShut
  {NoStop}%
\bibitem [{\citenamefont {Eichhorn}\ \emph {et~al.}(2009)\citenamefont
  {Eichhorn}, \citenamefont {Gies},\ and\ \citenamefont
  {Scherer}}]{Eichhorn:2009ah}%
  \BibitemOpen
  \bibfield  {author} {\bibinfo {author} {\bibfnamefont {A.}~\bibnamefont
  {Eichhorn}}, \bibinfo {author} {\bibfnamefont {H.}~\bibnamefont {Gies}}, \
  and\ \bibinfo {author} {\bibfnamefont {M.~M.}\ \bibnamefont {Scherer}},\
  }\href {\doibase 10.1103/PhysRevD.80.104003} {\bibfield  {journal} {\bibinfo
  {journal} {Phys. Rev.}\ }\textbf {\bibinfo {volume} {D80}},\ \bibinfo {pages}
  {104003} (\bibinfo {year} {2009})},\ \Eprint {http://arxiv.org/abs/0907.1828}
  {arXiv:0907.1828 [hep-th]} \BibitemShut {NoStop}%
\bibitem [{\citenamefont {Manrique}\ \emph
  {et~al.}(2011{\natexlab{a}})\citenamefont {Manrique}, \citenamefont
  {Rechenberger},\ and\ \citenamefont {Saueressig}}]{Manrique:2011jc}%
  \BibitemOpen
  \bibfield  {author} {\bibinfo {author} {\bibfnamefont {E.}~\bibnamefont
  {Manrique}}, \bibinfo {author} {\bibfnamefont {S.}~\bibnamefont
  {Rechenberger}}, \ and\ \bibinfo {author} {\bibfnamefont {F.}~\bibnamefont
  {Saueressig}},\ }\href {\doibase 10.1103/PhysRevLett.106.251302} {\bibfield
  {journal} {\bibinfo  {journal} {Phys.Rev.Lett.}\ }\textbf {\bibinfo {volume}
  {106}},\ \bibinfo {pages} {251302} (\bibinfo {year} {2011}{\natexlab{a}})},\
  \Eprint {http://arxiv.org/abs/1102.5012} {arXiv:1102.5012 [hep-th]}
  \BibitemShut {NoStop}%
\bibitem [{\citenamefont {Rechenberger}\ and\ \citenamefont
  {Saueressig}(2012)}]{Rechenberger:2012pm}%
  \BibitemOpen
  \bibfield  {author} {\bibinfo {author} {\bibfnamefont {S.}~\bibnamefont
  {Rechenberger}}\ and\ \bibinfo {author} {\bibfnamefont {F.}~\bibnamefont
  {Saueressig}},\ }\href {\doibase 10.1103/PhysRevD.86.024018} {\bibfield
  {journal} {\bibinfo  {journal} {Phys.Rev.}\ }\textbf {\bibinfo {volume}
  {D86}},\ \bibinfo {pages} {024018} (\bibinfo {year} {2012})},\ \Eprint
  {http://arxiv.org/abs/1206.0657} {arXiv:1206.0657 [hep-th]} \BibitemShut
  {NoStop}%
\bibitem [{\citenamefont {Codello}\ \emph {et~al.}(2014)\citenamefont
  {Codello}, \citenamefont {D'Odorico},\ and\ \citenamefont
  {Pagani}}]{Codello:2013fpa}%
  \BibitemOpen
  \bibfield  {author} {\bibinfo {author} {\bibfnamefont {A.}~\bibnamefont
  {Codello}}, \bibinfo {author} {\bibfnamefont {G.}~\bibnamefont {D'Odorico}},
  \ and\ \bibinfo {author} {\bibfnamefont {C.}~\bibnamefont {Pagani}},\ }\href
  {\doibase 10.1103/PhysRevD.89.081701} {\bibfield  {journal} {\bibinfo
  {journal} {Phys. Rev.}\ }\textbf {\bibinfo {volume} {D89}},\ \bibinfo {pages}
  {081701} (\bibinfo {year} {2014})},\ \Eprint {http://arxiv.org/abs/1304.4777}
  {arXiv:1304.4777 [gr-qc]} \BibitemShut {NoStop}%
\bibitem [{\citenamefont {Falls}\ \emph {et~al.}(2013)\citenamefont {Falls},
  \citenamefont {Litim}, \citenamefont {Nikolakopoulos},\ and\ \citenamefont
  {Rahmede}}]{Falls:2013bv}%
  \BibitemOpen
  \bibfield  {author} {\bibinfo {author} {\bibfnamefont {K.}~\bibnamefont
  {Falls}}, \bibinfo {author} {\bibfnamefont {D.}~\bibnamefont {Litim}},
  \bibinfo {author} {\bibfnamefont {K.}~\bibnamefont {Nikolakopoulos}}, \ and\
  \bibinfo {author} {\bibfnamefont {C.}~\bibnamefont {Rahmede}},\ }\href@noop
  {} {\  (\bibinfo {year} {2013})},\ \Eprint {http://arxiv.org/abs/1301.4191}
  {arXiv:1301.4191 [hep-th]} \BibitemShut {NoStop}%
\bibitem [{\citenamefont {Falls}(2016)}]{Falls:2014zba}%
  \BibitemOpen
  \bibfield  {author} {\bibinfo {author} {\bibfnamefont {K.}~\bibnamefont
  {Falls}},\ }\href {\doibase 10.1007/JHEP01(2016)069} {\bibfield  {journal}
  {\bibinfo  {journal} {JHEP}\ }\textbf {\bibinfo {volume} {01}},\ \bibinfo
  {pages} {069} (\bibinfo {year} {2016})},\ \Eprint
  {http://arxiv.org/abs/1408.0276} {arXiv:1408.0276 [hep-th]} \BibitemShut
  {NoStop}%
\bibitem [{\citenamefont {Falls}(2015)}]{Falls:2015qga}%
  \BibitemOpen
  \bibfield  {author} {\bibinfo {author} {\bibfnamefont {K.}~\bibnamefont
  {Falls}},\ }\href {\doibase 10.1103/PhysRevD.92.124057} {\bibfield  {journal}
  {\bibinfo  {journal} {Phys. Rev.}\ }\textbf {\bibinfo {volume} {D92}},\
  \bibinfo {pages} {124057} (\bibinfo {year} {2015})},\ \Eprint
  {http://arxiv.org/abs/1501.05331} {arXiv:1501.05331 [hep-th]} \BibitemShut
  {NoStop}%
\bibitem [{\citenamefont {Gies}\ \emph {et~al.}(2015)\citenamefont {Gies},
  \citenamefont {Knorr},\ and\ \citenamefont {Lippoldt}}]{Gies:2015tca}%
  \BibitemOpen
  \bibfield  {author} {\bibinfo {author} {\bibfnamefont {H.}~\bibnamefont
  {Gies}}, \bibinfo {author} {\bibfnamefont {B.}~\bibnamefont {Knorr}}, \ and\
  \bibinfo {author} {\bibfnamefont {S.}~\bibnamefont {Lippoldt}},\ }\href
  {\doibase 10.1103/PhysRevD.92.084020} {\bibfield  {journal} {\bibinfo
  {journal} {Phys. Rev.}\ }\textbf {\bibinfo {volume} {D92}},\ \bibinfo {pages}
  {084020} (\bibinfo {year} {2015})},\ \Eprint
  {http://arxiv.org/abs/1507.08859} {arXiv:1507.08859 [hep-th]} \BibitemShut
  {NoStop}%
\bibitem [{\citenamefont {Niedermaier}\ and\ \citenamefont
  {Reuter}(2006)}]{Niedermaier:2006wt}%
  \BibitemOpen
  \bibfield  {author} {\bibinfo {author} {\bibfnamefont {M.}~\bibnamefont
  {Niedermaier}}\ and\ \bibinfo {author} {\bibfnamefont {M.}~\bibnamefont
  {Reuter}},\ }\href@noop {} {\bibfield  {journal} {\bibinfo  {journal} {Living
  Rev.Rel.}\ }\textbf {\bibinfo {volume} {9}},\ \bibinfo {pages} {5} (\bibinfo
  {year} {2006})}\BibitemShut {NoStop}%
\bibitem [{\citenamefont {Percacci}(2007)}]{Percacci:2007sz}%
  \BibitemOpen
  \bibfield  {author} {\bibinfo {author} {\bibfnamefont {R.}~\bibnamefont
  {Percacci}},\ }\href@noop {} {\bibfield  {journal} {\bibinfo  {journal} {In
  *Oriti, D. (ed.): Approaches to quantum gravity* 111-128}\ } (\bibinfo {year}
  {2007})},\ \Eprint {http://arxiv.org/abs/0709.3851} {arXiv:0709.3851
  [hep-th]} \BibitemShut {NoStop}%
\bibitem [{\citenamefont {Litim}(2011)}]{Litim:2011cp}%
  \BibitemOpen
  \bibfield  {author} {\bibinfo {author} {\bibfnamefont {D.~F.}\ \bibnamefont
  {Litim}},\ }\href@noop {} {\bibfield  {journal} {\bibinfo  {journal}
  {Phil.Trans.Roy.Soc.Lond.}\ }\textbf {\bibinfo {volume} {A369}},\ \bibinfo
  {pages} {2759} (\bibinfo {year} {2011})},\ \Eprint
  {http://arxiv.org/abs/1102.4624} {arXiv:1102.4624 [hep-th]} \BibitemShut
  {NoStop}%
\bibitem [{\citenamefont {Reuter}\ and\ \citenamefont
  {Saueressig}(2012)}]{Reuter:2012id}%
  \BibitemOpen
  \bibfield  {author} {\bibinfo {author} {\bibfnamefont {M.}~\bibnamefont
  {Reuter}}\ and\ \bibinfo {author} {\bibfnamefont {F.}~\bibnamefont
  {Saueressig}},\ }\href {\doibase 10.1088/1367-2630/14/5/055022} {\bibfield
  {journal} {\bibinfo  {journal} {New J. Phys.}\ }\textbf {\bibinfo {volume}
  {14}},\ \bibinfo {pages} {055022} (\bibinfo {year} {2012})},\ \Eprint
  {http://arxiv.org/abs/1202.2274} {arXiv:1202.2274 [hep-th]} \BibitemShut
  {NoStop}%
\bibitem [{\citenamefont {Dou}\ and\ \citenamefont
  {Percacci}(1998)}]{Dou:1997fg}%
  \BibitemOpen
  \bibfield  {author} {\bibinfo {author} {\bibfnamefont {D.}~\bibnamefont
  {Dou}}\ and\ \bibinfo {author} {\bibfnamefont {R.}~\bibnamefont {Percacci}},\
  }\href {\doibase 10.1088/0264-9381/15/11/011} {\bibfield  {journal} {\bibinfo
   {journal} {Class. Quant. Grav.}\ }\textbf {\bibinfo {volume} {15}},\
  \bibinfo {pages} {3449} (\bibinfo {year} {1998})},\ \Eprint
  {http://arxiv.org/abs/hep-th/9707239} {arXiv:hep-th/9707239 [hep-th]}
  \BibitemShut {NoStop}%
\bibitem [{\citenamefont {Percacci}\ and\ \citenamefont
  {Perini}(2003{\natexlab{a}})}]{Percacci:2002ie}%
  \BibitemOpen
  \bibfield  {author} {\bibinfo {author} {\bibfnamefont {R.}~\bibnamefont
  {Percacci}}\ and\ \bibinfo {author} {\bibfnamefont {D.}~\bibnamefont
  {Perini}},\ }\href {\doibase 10.1103/PhysRevD.67.081503} {\bibfield
  {journal} {\bibinfo  {journal} {Phys. Rev.}\ }\textbf {\bibinfo {volume}
  {D67}},\ \bibinfo {pages} {081503} (\bibinfo {year} {2003}{\natexlab{a}})},\
  \Eprint {http://arxiv.org/abs/hep-th/0207033} {arXiv:hep-th/0207033}
  \BibitemShut {NoStop}%
\bibitem [{\citenamefont {Percacci}\ and\ \citenamefont
  {Perini}(2003{\natexlab{b}})}]{Percacci:2003jz}%
  \BibitemOpen
  \bibfield  {author} {\bibinfo {author} {\bibfnamefont {R.}~\bibnamefont
  {Percacci}}\ and\ \bibinfo {author} {\bibfnamefont {D.}~\bibnamefont
  {Perini}},\ }\href {\doibase 10.1103/PhysRevD.68.044018} {\bibfield
  {journal} {\bibinfo  {journal} {Phys. Rev.}\ }\textbf {\bibinfo {volume}
  {D68}},\ \bibinfo {pages} {044018} (\bibinfo {year} {2003}{\natexlab{b}})},\
  \Eprint {http://arxiv.org/abs/hep-th/0304222} {arXiv:hep-th/0304222}
  \BibitemShut {NoStop}%
\bibitem [{\citenamefont {Folkerts}\ \emph {et~al.}(2012)\citenamefont
  {Folkerts}, \citenamefont {Litim},\ and\ \citenamefont
  {Pawlowski}}]{Folkerts:2011jz}%
  \BibitemOpen
  \bibfield  {author} {\bibinfo {author} {\bibfnamefont {S.}~\bibnamefont
  {Folkerts}}, \bibinfo {author} {\bibfnamefont {D.~F.}\ \bibnamefont {Litim}},
  \ and\ \bibinfo {author} {\bibfnamefont {J.~M.}\ \bibnamefont {Pawlowski}},\
  }\href {\doibase 10.1016/j.physletb.2012.02.002} {\bibfield  {journal}
  {\bibinfo  {journal} {Phys.Lett.}\ }\textbf {\bibinfo {volume} {B709}},\
  \bibinfo {pages} {234} (\bibinfo {year} {2012})},\ \Eprint
  {http://arxiv.org/abs/1101.5552} {arXiv:1101.5552 [hep-th]} \BibitemShut
  {NoStop}%
\bibitem [{\citenamefont {Don{\`a}}\ and\ \citenamefont
  {Percacci}(2013)}]{Dona:2012am}%
  \BibitemOpen
  \bibfield  {author} {\bibinfo {author} {\bibfnamefont {P.}~\bibnamefont
  {Don{\`a}}}\ and\ \bibinfo {author} {\bibfnamefont {R.}~\bibnamefont
  {Percacci}},\ }\href {\doibase 10.1103/PhysRevD.87.045002} {\bibfield
  {journal} {\bibinfo  {journal} {Phys. Rev.}\ }\textbf {\bibinfo {volume}
  {D87}},\ \bibinfo {pages} {045002} (\bibinfo {year} {2013})},\ \Eprint
  {http://arxiv.org/abs/1209.3649} {arXiv:1209.3649 [hep-th]} \BibitemShut
  {NoStop}%
\bibitem [{\citenamefont {Don{\`a}}\ \emph {et~al.}(2014)\citenamefont
  {Don{\`a}}, \citenamefont {Eichhorn},\ and\ \citenamefont
  {Percacci}}]{Dona:2013qba}%
  \BibitemOpen
  \bibfield  {author} {\bibinfo {author} {\bibfnamefont {P.}~\bibnamefont
  {Don{\`a}}}, \bibinfo {author} {\bibfnamefont {A.}~\bibnamefont {Eichhorn}},
  \ and\ \bibinfo {author} {\bibfnamefont {R.}~\bibnamefont {Percacci}},\
  }\href {\doibase 10.1103/PhysRevD.89.084035} {\bibfield  {journal} {\bibinfo
  {journal} {Phys.Rev.}\ }\textbf {\bibinfo {volume} {D89}},\ \bibinfo {pages}
  {084035} (\bibinfo {year} {2014})},\ \Eprint {http://arxiv.org/abs/1311.2898}
  {arXiv:1311.2898 [hep-th]} \BibitemShut {NoStop}%
\bibitem [{\citenamefont {Don{\`a}}\ \emph {et~al.}(2015)\citenamefont
  {Don{\`a}}, \citenamefont {Eichhorn},\ and\ \citenamefont
  {Percacci}}]{Dona:2014pla}%
  \BibitemOpen
  \bibfield  {author} {\bibinfo {author} {\bibfnamefont {P.}~\bibnamefont
  {Don{\`a}}}, \bibinfo {author} {\bibfnamefont {A.}~\bibnamefont {Eichhorn}},
  \ and\ \bibinfo {author} {\bibfnamefont {R.}~\bibnamefont {Percacci}},\
  }\bibfield  {booktitle} {\emph {\bibinfo {booktitle} {{Proceedings, Satellite
  Conference on Theory Canada 9}}},\ }\href {\doibase 10.1139/cjp-2014-0574}
  {\bibfield  {journal} {\bibinfo  {journal} {Can. J. Phys.}\ }\textbf
  {\bibinfo {volume} {93}},\ \bibinfo {pages} {988} (\bibinfo {year} {2015})},\
  \Eprint {http://arxiv.org/abs/1410.4411} {arXiv:1410.4411 [gr-qc]}
  \BibitemShut {NoStop}%
\bibitem [{\citenamefont {Oda}\ and\ \citenamefont
  {Yamada}(2015)}]{Oda:2015sma}%
  \BibitemOpen
  \bibfield  {author} {\bibinfo {author} {\bibfnamefont {K.-y.}\ \bibnamefont
  {Oda}}\ and\ \bibinfo {author} {\bibfnamefont {M.}~\bibnamefont {Yamada}},\
  }\href@noop {} {\  (\bibinfo {year} {2015})},\ \Eprint
  {http://arxiv.org/abs/1510.03734} {arXiv:1510.03734 [hep-th]} \BibitemShut
  {NoStop}%
\bibitem [{\citenamefont {Pawlowski}(2002)}]{Pawlowski:2002eb}%
  \BibitemOpen
  \bibfield  {author} {\bibinfo {author} {\bibfnamefont {J.~M.}\ \bibnamefont
  {Pawlowski}},\ }\href@noop {} {\bibfield  {journal} {\bibinfo  {journal}
  {Acta Phys.Slov.}\ }\textbf {\bibinfo {volume} {52}},\ \bibinfo {pages} {475}
  (\bibinfo {year} {2002})}\BibitemShut {NoStop}%
\bibitem [{\citenamefont {Litim}\ and\ \citenamefont
  {Pawlowski}(2002{\natexlab{a}})}]{Litim:2002ce}%
  \BibitemOpen
  \bibfield  {author} {\bibinfo {author} {\bibfnamefont {D.~F.}\ \bibnamefont
  {Litim}}\ and\ \bibinfo {author} {\bibfnamefont {J.~M.}\ \bibnamefont
  {Pawlowski}},\ }\href@noop {} {\bibfield  {journal} {\bibinfo  {journal}
  {JHEP}\ }\textbf {\bibinfo {volume} {0209}},\ \bibinfo {pages} {049}
  (\bibinfo {year} {2002}{\natexlab{a}})},\ \Eprint
  {http://arxiv.org/abs/hep-th/0203005} {arXiv:hep-th/0203005 [hep-th]}
  \BibitemShut {NoStop}%
\bibitem [{\citenamefont {Litim}\ and\ \citenamefont
  {Pawlowski}(2002{\natexlab{b}})}]{Litim:2002hj}%
  \BibitemOpen
  \bibfield  {author} {\bibinfo {author} {\bibfnamefont {D.~F.}\ \bibnamefont
  {Litim}}\ and\ \bibinfo {author} {\bibfnamefont {J.~M.}\ \bibnamefont
  {Pawlowski}},\ }\href {\doibase 10.1016/S0370-2693(02)02693-X} {\bibfield
  {journal} {\bibinfo  {journal} {Phys.Lett.}\ }\textbf {\bibinfo {volume}
  {B546}},\ \bibinfo {pages} {279} (\bibinfo {year} {2002}{\natexlab{b}})},\
  \Eprint {http://arxiv.org/abs/hep-th/0208216} {arXiv:hep-th/0208216 [hep-th]}
  \BibitemShut {NoStop}%
\bibitem [{\citenamefont {Pawlowski}(2003)}]{Pawlowski:2003sk}%
  \BibitemOpen
  \bibfield  {author} {\bibinfo {author} {\bibfnamefont {J.~M.}\ \bibnamefont
  {Pawlowski}},\ }\href@noop {} {\  (\bibinfo {year} {2003})},\ \Eprint
  {http://arxiv.org/abs/hep-th/0310018} {arXiv:hep-th/0310018 [hep-th]}
  \BibitemShut {NoStop}%
\bibitem [{\citenamefont {Pawlowski}(2007)}]{Pawlowski:2005xe}%
  \BibitemOpen
  \bibfield  {author} {\bibinfo {author} {\bibfnamefont {J.~M.}\ \bibnamefont
  {Pawlowski}},\ }\href {\doibase 10.1016/j.aop.2007.01.007} {\bibfield
  {journal} {\bibinfo  {journal} {Annals Phys.}\ }\textbf {\bibinfo {volume}
  {322}},\ \bibinfo {pages} {2831} (\bibinfo {year} {2007})},\ \Eprint
  {http://arxiv.org/abs/hep-th/0512261} {arXiv:hep-th/0512261 [hep-th]}
  \BibitemShut {NoStop}%
\bibitem [{\citenamefont {Branchina}\ \emph {et~al.}(2003)\citenamefont
  {Branchina}, \citenamefont {Meissner},\ and\ \citenamefont
  {Veneziano}}]{Branchina:2003ek}%
  \BibitemOpen
  \bibfield  {author} {\bibinfo {author} {\bibfnamefont {V.}~\bibnamefont
  {Branchina}}, \bibinfo {author} {\bibfnamefont {K.~A.}\ \bibnamefont
  {Meissner}}, \ and\ \bibinfo {author} {\bibfnamefont {G.}~\bibnamefont
  {Veneziano}},\ }\href {\doibase 10.1016/j.physletb.2003.09.020} {\bibfield
  {journal} {\bibinfo  {journal} {Phys.Lett.}\ }\textbf {\bibinfo {volume}
  {B574}},\ \bibinfo {pages} {319} (\bibinfo {year} {2003})},\ \Eprint
  {http://arxiv.org/abs/hep-th/0309234} {arXiv:hep-th/0309234 [hep-th]}
  \BibitemShut {NoStop}%
\bibitem [{\citenamefont {Demmel}\ \emph {et~al.}(2015)\citenamefont {Demmel},
  \citenamefont {Saueressig},\ and\ \citenamefont {Zanusso}}]{Demmel:2014hla}%
  \BibitemOpen
  \bibfield  {author} {\bibinfo {author} {\bibfnamefont {M.}~\bibnamefont
  {Demmel}}, \bibinfo {author} {\bibfnamefont {F.}~\bibnamefont {Saueressig}},
  \ and\ \bibinfo {author} {\bibfnamefont {O.}~\bibnamefont {Zanusso}},\ }\href
  {\doibase 10.1016/j.aop.2015.04.018} {\bibfield  {journal} {\bibinfo
  {journal} {Annals Phys.}\ }\textbf {\bibinfo {volume} {359}},\ \bibinfo
  {pages} {141} (\bibinfo {year} {2015})},\ \Eprint
  {http://arxiv.org/abs/1412.7207} {arXiv:1412.7207 [hep-th]} \BibitemShut
  {NoStop}%
\bibitem [{\citenamefont {Demmel}\ and\ \citenamefont
  {Nink}(2015)}]{Demmel:2015zfa}%
  \BibitemOpen
  \bibfield  {author} {\bibinfo {author} {\bibfnamefont {M.}~\bibnamefont
  {Demmel}}\ and\ \bibinfo {author} {\bibfnamefont {A.}~\bibnamefont {Nink}},\
  }\href {\doibase 10.1103/PhysRevD.92.104013} {\bibfield  {journal} {\bibinfo
  {journal} {Phys. Rev.}\ }\textbf {\bibinfo {volume} {D92}},\ \bibinfo {pages}
  {104013} (\bibinfo {year} {2015})},\ \Eprint
  {http://arxiv.org/abs/1506.03809} {arXiv:1506.03809 [gr-qc]} \BibitemShut
  {NoStop}%
\bibitem [{\citenamefont {Safari}(2015)}]{Safari:2015dva}%
  \BibitemOpen
  \bibfield  {author} {\bibinfo {author} {\bibfnamefont {M.}~\bibnamefont
  {Safari}},\ }\href@noop {} {\  (\bibinfo {year} {2015})},\ \Eprint
  {http://arxiv.org/abs/1508.06244} {arXiv:1508.06244 [hep-th]} \BibitemShut
  {NoStop}%
\bibitem [{\citenamefont {Manrique}\ and\ \citenamefont
  {Reuter}(2010)}]{Manrique:2009uh}%
  \BibitemOpen
  \bibfield  {author} {\bibinfo {author} {\bibfnamefont {E.}~\bibnamefont
  {Manrique}}\ and\ \bibinfo {author} {\bibfnamefont {M.}~\bibnamefont
  {Reuter}},\ }\href {\doibase 10.1016/j.aop.2009.11.009} {\bibfield  {journal}
  {\bibinfo  {journal} {Annals Phys.}\ }\textbf {\bibinfo {volume} {325}},\
  \bibinfo {pages} {785} (\bibinfo {year} {2010})},\ \Eprint
  {http://arxiv.org/abs/0907.2617} {arXiv:0907.2617 [gr-qc]} \BibitemShut
  {NoStop}%
\bibitem [{\citenamefont {Manrique}\ \emph
  {et~al.}(2011{\natexlab{b}})\citenamefont {Manrique}, \citenamefont
  {Reuter},\ and\ \citenamefont {Saueressig}}]{Manrique:2010am}%
  \BibitemOpen
  \bibfield  {author} {\bibinfo {author} {\bibfnamefont {E.}~\bibnamefont
  {Manrique}}, \bibinfo {author} {\bibfnamefont {M.}~\bibnamefont {Reuter}}, \
  and\ \bibinfo {author} {\bibfnamefont {F.}~\bibnamefont {Saueressig}},\
  }\href {\doibase 10.1016/j.aop.2010.11.006} {\bibfield  {journal} {\bibinfo
  {journal} {Annals Phys.}\ }\textbf {\bibinfo {volume} {326}},\ \bibinfo
  {pages} {463} (\bibinfo {year} {2011}{\natexlab{b}})},\ \Eprint
  {http://arxiv.org/abs/1006.0099} {arXiv:1006.0099 [hep-th]} \BibitemShut
  {NoStop}%
\bibitem [{\citenamefont {Reuter}\ and\ \citenamefont
  {Wetterich}(1994)}]{Reuter:1993kw}%
  \BibitemOpen
  \bibfield  {author} {\bibinfo {author} {\bibfnamefont {M.}~\bibnamefont
  {Reuter}}\ and\ \bibinfo {author} {\bibfnamefont {C.}~\bibnamefont
  {Wetterich}},\ }\href {\doibase 10.1016/0550-3213(94)90543-6} {\bibfield
  {journal} {\bibinfo  {journal} {Nucl. Phys.}\ }\textbf {\bibinfo {volume}
  {B417}},\ \bibinfo {pages} {181} (\bibinfo {year} {1994})}\BibitemShut
  {NoStop}%
\bibitem [{\citenamefont {Ellwanger}(1994)}]{Ellwanger:1993mw}%
  \BibitemOpen
  \bibfield  {author} {\bibinfo {author} {\bibfnamefont {U.}~\bibnamefont
  {Ellwanger}},\ }\href {\doibase 10.1007/BF01555911} {\bibfield  {journal}
  {\bibinfo  {journal} {Z. Phys.}\ }\textbf {\bibinfo {volume} {C62}},\
  \bibinfo {pages} {503} (\bibinfo {year} {1994})},\ \Eprint
  {http://arxiv.org/abs/hep-ph/9308260} {arXiv:hep-ph/9308260 [hep-ph]}
  \BibitemShut {NoStop}%
\bibitem [{\citenamefont {Morris}(1994)}]{Morris:1993qb}%
  \BibitemOpen
  \bibfield  {author} {\bibinfo {author} {\bibfnamefont {T.~R.}\ \bibnamefont
  {Morris}},\ }\href {\doibase 10.1142/S0217751X94000972} {\bibfield  {journal}
  {\bibinfo  {journal} {Int. J. Mod. Phys.}\ }\textbf {\bibinfo {volume}
  {A9}},\ \bibinfo {pages} {2411} (\bibinfo {year} {1994})},\ \Eprint
  {http://arxiv.org/abs/hep-ph/9308265} {arXiv:hep-ph/9308265} \BibitemShut
  {NoStop}%
\bibitem [{\citenamefont {Eichhorn}\ and\ \citenamefont
  {Gies}(2010)}]{Eichhorn:2010tb}%
  \BibitemOpen
  \bibfield  {author} {\bibinfo {author} {\bibfnamefont {A.}~\bibnamefont
  {Eichhorn}}\ and\ \bibinfo {author} {\bibfnamefont {H.}~\bibnamefont
  {Gies}},\ }\href {\doibase 10.1103/PhysRevD.81.104010} {\bibfield  {journal}
  {\bibinfo  {journal} {Phys. Rev.}\ }\textbf {\bibinfo {volume} {D81}},\
  \bibinfo {pages} {104010} (\bibinfo {year} {2010})},\ \Eprint
  {http://arxiv.org/abs/1001.5033} {arXiv:1001.5033 [hep-th]} \BibitemShut
  {NoStop}%
\bibitem [{\citenamefont {Fischer}\ and\ \citenamefont
  {Pawlowski}(2009)}]{Fischer:2009tn}%
  \BibitemOpen
  \bibfield  {author} {\bibinfo {author} {\bibfnamefont {C.~S.}\ \bibnamefont
  {Fischer}}\ and\ \bibinfo {author} {\bibfnamefont {J.~M.}\ \bibnamefont
  {Pawlowski}},\ }\href {\doibase 10.1103/PhysRevD.80.025023} {\bibfield
  {journal} {\bibinfo  {journal} {Phys. Rev.}\ }\textbf {\bibinfo {volume}
  {D80}},\ \bibinfo {pages} {025023} (\bibinfo {year} {2009})},\ \Eprint
  {http://arxiv.org/abs/0903.2193} {arXiv:0903.2193 [hep-th]} \BibitemShut
  {NoStop}%
\bibitem [{\citenamefont {Weldon}(2001)}]{Weldon:2000fr}%
  \BibitemOpen
  \bibfield  {author} {\bibinfo {author} {\bibfnamefont {H.~A.}\ \bibnamefont
  {Weldon}},\ }\href {\doibase 10.1103/PhysRevD.63.104010} {\bibfield
  {journal} {\bibinfo  {journal} {Phys. Rev.}\ }\textbf {\bibinfo {volume}
  {D63}},\ \bibinfo {pages} {104010} (\bibinfo {year} {2001})},\ \Eprint
  {http://arxiv.org/abs/gr-qc/0009086} {arXiv:gr-qc/0009086 [gr-qc]}
  \BibitemShut {NoStop}%
\bibitem [{\citenamefont {Gies}\ and\ \citenamefont
  {Lippoldt}(2014)}]{Gies:2013noa}%
  \BibitemOpen
  \bibfield  {author} {\bibinfo {author} {\bibfnamefont {H.}~\bibnamefont
  {Gies}}\ and\ \bibinfo {author} {\bibfnamefont {S.}~\bibnamefont
  {Lippoldt}},\ }\href {\doibase 10.1103/PhysRevD.89.064040} {\bibfield
  {journal} {\bibinfo  {journal} {Phys.Rev.}\ }\textbf {\bibinfo {volume}
  {D89}},\ \bibinfo {pages} {064040} (\bibinfo {year} {2014})},\ \Eprint
  {http://arxiv.org/abs/1310.2509} {arXiv:1310.2509 [hep-th]} \BibitemShut
  {NoStop}%
\bibitem [{\citenamefont {Lippoldt}(2015)}]{Lippoldt:2015cea}%
  \BibitemOpen
  \bibfield  {author} {\bibinfo {author} {\bibfnamefont {S.}~\bibnamefont
  {Lippoldt}},\ }\href {\doibase 10.1103/PhysRevD.91.104006} {\bibfield
  {journal} {\bibinfo  {journal} {Phys. Rev.}\ }\textbf {\bibinfo {volume}
  {D91}},\ \bibinfo {pages} {104006} (\bibinfo {year} {2015})},\ \Eprint
  {http://arxiv.org/abs/1502.05607} {arXiv:1502.05607 [hep-th]} \BibitemShut
  {NoStop}%
\bibitem [{\citenamefont {Litim}\ and\ \citenamefont
  {Pawlowski}(1998)}]{Litim:1998qi}%
  \BibitemOpen
  \bibfield  {author} {\bibinfo {author} {\bibfnamefont {D.~F.}\ \bibnamefont
  {Litim}}\ and\ \bibinfo {author} {\bibfnamefont {J.~M.}\ \bibnamefont
  {Pawlowski}},\ }\href {\doibase 10.1016/S0370-2693(98)00761-8} {\bibfield
  {journal} {\bibinfo  {journal} {Phys.Lett.}\ }\textbf {\bibinfo {volume}
  {B435}},\ \bibinfo {pages} {181} (\bibinfo {year} {1998})},\ \Eprint
  {http://arxiv.org/abs/hep-th/9802064} {arXiv:hep-th/9802064 [hep-th]}
  \BibitemShut {NoStop}%
\bibitem [{\citenamefont {Kuipers}\ \emph {et~al.}(2013)\citenamefont
  {Kuipers}, \citenamefont {Ueda}, \citenamefont {Vermaseren},\ and\
  \citenamefont {Vollinga}}]{Kuipers:2012rf}%
  \BibitemOpen
  \bibfield  {author} {\bibinfo {author} {\bibfnamefont {J.}~\bibnamefont
  {Kuipers}}, \bibinfo {author} {\bibfnamefont {T.}~\bibnamefont {Ueda}},
  \bibinfo {author} {\bibfnamefont {J.~A.~M.}\ \bibnamefont {Vermaseren}}, \
  and\ \bibinfo {author} {\bibfnamefont {J.}~\bibnamefont {Vollinga}},\ }\href
  {\doibase 10.1016/j.cpc.2012.12.028} {\bibfield  {journal} {\bibinfo
  {journal} {Comput. Phys. Commun.}\ }\textbf {\bibinfo {volume} {184}},\
  \bibinfo {pages} {1453} (\bibinfo {year} {2013})},\ \Eprint
  {http://arxiv.org/abs/1203.6543} {arXiv:1203.6543 [cs.SC]} \BibitemShut
  {NoStop}%
\bibitem [{\citenamefont {Vermaseren}(2000)}]{Vermaseren:2000nd}%
  \BibitemOpen
  \bibfield  {author} {\bibinfo {author} {\bibfnamefont {J.~A.~M.}\
  \bibnamefont {Vermaseren}},\ }\href@noop {} {\  (\bibinfo {year} {2000})},\
  \Eprint {http://arxiv.org/abs/math-ph/0010025} {arXiv:math-ph/0010025
  [math-ph]} \BibitemShut {NoStop}%
\bibitem [{\citenamefont {Groh}\ and\ \citenamefont
  {Saueressig}(2010)}]{Groh:2010ta}%
  \BibitemOpen
  \bibfield  {author} {\bibinfo {author} {\bibfnamefont {K.}~\bibnamefont
  {Groh}}\ and\ \bibinfo {author} {\bibfnamefont {F.}~\bibnamefont
  {Saueressig}},\ }\href {\doibase 10.1088/1751-8113/43/36/365403} {\bibfield
  {journal} {\bibinfo  {journal} {J. Phys.}\ }\textbf {\bibinfo {volume}
  {A43}},\ \bibinfo {pages} {365403} (\bibinfo {year} {2010})},\ \Eprint
  {http://arxiv.org/abs/1001.5032} {arXiv:1001.5032 [hep-th]} \BibitemShut
  {NoStop}%
\bibitem [{\citenamefont {Pawlowski}(2001)}]{Pawlowski:2001df}%
  \BibitemOpen
  \bibfield  {author} {\bibinfo {author} {\bibfnamefont {J.~M.}\ \bibnamefont
  {Pawlowski}},\ }\href {\doibase 10.1142/S0217751X01004785} {\bibfield
  {journal} {\bibinfo  {journal} {Int.J.Mod.Phys.}\ }\textbf {\bibinfo {volume}
  {A16}},\ \bibinfo {pages} {2105} (\bibinfo {year} {2001})}\BibitemShut
  {NoStop}%
\bibitem [{\citenamefont {Gies}(2002)}]{Gies:2002af}%
  \BibitemOpen
  \bibfield  {author} {\bibinfo {author} {\bibfnamefont {H.}~\bibnamefont
  {Gies}},\ }\href {\doibase 10.1103/PhysRevD.66.025006} {\bibfield  {journal}
  {\bibinfo  {journal} {Phys. Rev.}\ }\textbf {\bibinfo {volume} {D66}},\
  \bibinfo {pages} {025006} (\bibinfo {year} {2002})},\ \Eprint
  {http://arxiv.org/abs/hep-th/0202207} {arXiv:hep-th/0202207 [hep-th]}
  \BibitemShut {NoStop}%
\bibitem [{\citenamefont {Litim}(2000)}]{Litim:2000ci}%
  \BibitemOpen
  \bibfield  {author} {\bibinfo {author} {\bibfnamefont {D.~F.}\ \bibnamefont
  {Litim}},\ }\href {\doibase 10.1016/S0370-2693(00)00748-6} {\bibfield
  {journal} {\bibinfo  {journal} {Phys.Lett.}\ }\textbf {\bibinfo {volume}
  {B486}},\ \bibinfo {pages} {92} (\bibinfo {year} {2000})},\ \Eprint
  {http://arxiv.org/abs/hep-th/0005245} {arXiv:hep-th/0005245 [hep-th]}
  \BibitemShut {NoStop}%
\bibitem [{\citenamefont {Eichhorn}\ and\ \citenamefont
  {Gies}(2011)}]{Eichhorn:2011pc}%
  \BibitemOpen
  \bibfield  {author} {\bibinfo {author} {\bibfnamefont {A.}~\bibnamefont
  {Eichhorn}}\ and\ \bibinfo {author} {\bibfnamefont {H.}~\bibnamefont
  {Gies}},\ }\href {\doibase 10.1088/1367-2630/13/12/125012} {\bibfield
  {journal} {\bibinfo  {journal} {New J. Phys.}\ }\textbf {\bibinfo {volume}
  {13}},\ \bibinfo {pages} {125012} (\bibinfo {year} {2011})},\ \Eprint
  {http://arxiv.org/abs/1104.5366} {arXiv:1104.5366 [hep-th]} \BibitemShut
  {NoStop}%
\bibitem [{\citenamefont {Eichhorn}(2012)}]{Eichhorn:2012va}%
  \BibitemOpen
  \bibfield  {author} {\bibinfo {author} {\bibfnamefont {A.}~\bibnamefont
  {Eichhorn}},\ }\href {\doibase 10.1103/PhysRevD.86.105021} {\bibfield
  {journal} {\bibinfo  {journal} {Phys. Rev.}\ }\textbf {\bibinfo {volume}
  {D86}},\ \bibinfo {pages} {105021} (\bibinfo {year} {2012})},\ \Eprint
  {http://arxiv.org/abs/1204.0965} {arXiv:1204.0965 [gr-qc]} \BibitemShut
  {NoStop}%
\bibitem [{\citenamefont {Henz}\ \emph {et~al.}(2013)\citenamefont {Henz},
  \citenamefont {Pawlowski}, \citenamefont {Rodigast},\ and\ \citenamefont
  {Wetterich}}]{Henz:2013oxa}%
  \BibitemOpen
  \bibfield  {author} {\bibinfo {author} {\bibfnamefont {T.}~\bibnamefont
  {Henz}}, \bibinfo {author} {\bibfnamefont {J.~M.}\ \bibnamefont {Pawlowski}},
  \bibinfo {author} {\bibfnamefont {A.}~\bibnamefont {Rodigast}}, \ and\
  \bibinfo {author} {\bibfnamefont {C.}~\bibnamefont {Wetterich}},\ }\href
  {\doibase 10.1016/j.physletb.2013.10.015} {\bibfield  {journal} {\bibinfo
  {journal} {Phys. Lett.}\ }\textbf {\bibinfo {volume} {B727}},\ \bibinfo
  {pages} {298} (\bibinfo {year} {2013})},\ \Eprint
  {http://arxiv.org/abs/1304.7743} {arXiv:1304.7743 [hep-th]} \BibitemShut
  {NoStop}%
\bibitem [{\citenamefont {Bridle}\ \emph {et~al.}(2014)\citenamefont {Bridle},
  \citenamefont {Dietz},\ and\ \citenamefont {Morris}}]{Bridle:2013sra}%
  \BibitemOpen
  \bibfield  {author} {\bibinfo {author} {\bibfnamefont {I.~H.}\ \bibnamefont
  {Bridle}}, \bibinfo {author} {\bibfnamefont {J.~A.}\ \bibnamefont {Dietz}}, \
  and\ \bibinfo {author} {\bibfnamefont {T.~R.}\ \bibnamefont {Morris}},\
  }\href {\doibase 10.1007/JHEP03(2014)093} {\bibfield  {journal} {\bibinfo
  {journal} {JHEP}\ }\textbf {\bibinfo {volume} {03}},\ \bibinfo {pages} {093}
  (\bibinfo {year} {2014})},\ \Eprint {http://arxiv.org/abs/1312.2846}
  {arXiv:1312.2846 [hep-th]} \BibitemShut {NoStop}%
\bibitem [{\citenamefont {Dietz}\ and\ \citenamefont
  {Morris}(2015)}]{Dietz:2015owa}%
  \BibitemOpen
  \bibfield  {author} {\bibinfo {author} {\bibfnamefont {J.~A.}\ \bibnamefont
  {Dietz}}\ and\ \bibinfo {author} {\bibfnamefont {T.~R.}\ \bibnamefont
  {Morris}},\ }\href {\doibase 10.1007/JHEP04(2015)118} {\bibfield  {journal}
  {\bibinfo  {journal} {JHEP}\ }\textbf {\bibinfo {volume} {04}},\ \bibinfo
  {pages} {118} (\bibinfo {year} {2015})},\ \Eprint
  {http://arxiv.org/abs/1502.07396} {arXiv:1502.07396 [hep-th]} \BibitemShut
  {NoStop}%
\bibitem [{\citenamefont {Vilkovisky}(1984)}]{Vilkovisky:1984st}%
  \BibitemOpen
  \bibfield  {author} {\bibinfo {author} {\bibfnamefont {G.}~\bibnamefont
  {Vilkovisky}},\ }\href {\doibase 10.1016/0550-3213(84)90228-1} {\bibfield
  {journal} {\bibinfo  {journal} {Nucl.Phys.}\ }\textbf {\bibinfo {volume}
  {B234}},\ \bibinfo {pages} {125} (\bibinfo {year} {1984})}\BibitemShut
  {NoStop}%
\bibitem [{\citenamefont {DeWitt}(1988)}]{DeWitt:1988dq}%
  \BibitemOpen
  \bibfield  {author} {\bibinfo {author} {\bibfnamefont {B.~S.}\ \bibnamefont
  {DeWitt}},\ }\href@noop {} {\bibfield  {journal} {\bibinfo  {journal}
  {Quantum Field Theory and Quantum Statistics, Vol. 1, Batalin, I.A. (Ed.) et
  al.}\ ,\ \bibinfo {pages} {191}} (\bibinfo {year} {1988})}\BibitemShut
  {NoStop}%
\end{thebibliography}%

\end{document}